\documentclass[12pt,titlepage,a4paper]{article}
\usepackage{amssymb}
\usepackage{amsmath}
\usepackage[mathscr]{eucal}

\advance\textwidth2cm
\advance\textheight3cm
\advance\oddsidemargin-1.0cm
\advance\topmargin-2cm

\newcommand{\be}{\begin{equation}}
\newcommand{\ee}{\end{equation}}
\newcommand{\bea}{\begin{eqnarray}}
\newcommand{\eea}{\end{eqnarray}}
\newcommand{\beas}{\begin{eqnarray*}}
\newcommand{\eeas}{\end{eqnarray*}}

\newcommand{\bs}{\begin{sloppypar}}
\newcommand{\es}{\end{sloppypar}}

\newcommand{\ie}{{\it i.e.}}
\newcommand{\eg}{{\it e.g.}}
\newcommand{\cf}{{\it cf.}}

\newcommand{\N}{\mathbb{N}}
\newcommand{\Z}{\mathbb{Z}}

\newcommand{\imag}{{i}}

\newcommand{\pa}{\partial}

\newcommand{\PE}[2]{P_{1|#2}^{[#1]}}
\newcommand{\PN}[2]{(P_{1|#2}^{[#1] 2})_{0}}
\newcommand{\PM}[3]{(P_{1|#3}^{[#2] #1})_{#1}}

\newcommand{\PZ}[1]{P_{0|2}^{[#1]}}
\newcommand{\PO}[1]{P_{0|0}^{[#1]}}
\newcommand{\poe}{P_{0|0}^{[1]}}
\newcommand{\poz}{P_{0|0}^{[2]}}
\newcommand{\pon}{P_{0|0}^{[0]}}
\newcommand{\pominus}{( \poe - \poz \;\!)}
\newcommand{\potimes}{( \poe\cdot\poz \;\!)}
\newcommand{\potimesq}{( \poe\cdot\poz \;\!)^2}

\newcommand{\mz}{\mathfrak{m}^{[2]}}
\newcommand{\mn}{\mathfrak{m}^{[0]}}
\newcommand{\mnq}{\mathfrak{m}^{[0]2}}

\newcommand{\XZQ}[1]{{\mathscr{X}_2^{[#1]}}}

\newcommand{\ZZQ}[1]{{\mathscr{Z}_2^{[#1]}}}

\newcommand{\XZC}[1]{{{X}_2^{[#1]}}}

\newcommand{\ZZC}[1]{{{Z}_2^{[#1]}}}

\newcommand{\JE}[1]{{J_1^{[#1]}}}
\newcommand{\JEQN}[1]{(J_1^{[#1]2})_{0}}
\newcommand{\JEQZ}[1]{(J_1^{[#1]2})_{2}}
\newcommand{\WE}[2]{W_{1|#2}^{[#1]}}
\renewcommand{\AE}{{\mathscr{A}}_{1|0}}
\newcommand{\BE}{{\mathscr{B}}_{1|0}}

\newcommand{\BNE}[1]{B_{0}^{(1)[#1]}}
\newcommand{\TZ}[1]{T_2^{[#1]}}
\newcommand{\SZ}[1]{S_2^{[#1]}}
\newcommand{\SE}[1]{S_1^{[#1]}}

\newcommand{\se}{s_1}
\newcommand{\te}{t_1}

\newcommand{\KN}{{\mathscr{K}}_0}
\newcommand{\KE}{{\mathscr{K}}_1}
\newcommand{\KZ}{{\mathscr{K}}_2}
\newcommand{\CN}{{\mathscr{C}}_0}
\newcommand{\CE}{{\mathscr{C}}_1}
\newcommand{\CZ}{{\mathscr{C}}_2}

\newcommand{\SO}{\mathscr{O}}
\newcommand{\SW}{\mathscr{W}}

\newcommand{\GV}{\mathfrak{V}}

\newcommand{\gh}{\mathfrak{h}}
\newcommand{\gm}{\mathfrak{m}}

\newcommand{\hq}{{\sf h}}
\begin{document}

\begin{titlepage}

\vspace*{-2cm}
\begin{flushright} { THEP 02/06\\University of Freiburg\\7th June 2002
}
\end{flushright}
\vspace{2cm}

\begin{center}
{\large\bf 
The Invariant Charges of the Nambu--Goto String Theory: \\[6mm]
Quantization of Non-Additive Composition Laws }\\[2.5cm]
{ K. Pohlmeyer and M. Trunk}\\[4mm]
{Fakult\"at f\"ur Physik der Universit\"at Freiburg,
Hermann--Herder--Str.\ 3, \\[3mm]
D--79104 Freiburg,
Germany}
\end{center}

\vspace{4cm}
\noindent
{\bf Abstract}:
We examine and implement the concept of non-additive composition laws
in the quantum theory of closed bosonic strings moving in
$(3+1)$-dimensional Minkowski space. Such laws supply exact selection
rules for the merging and splitting of closed strings.

\end{titlepage}

\section*{Introduction}

A relevant part of the quantum algebra formed by the infinitesimal
generators of observable symmetry transformations for the massive 
relativistic
closed bosonic string moving in (3+1)--dimensional space-time has
been constructed and analyzed Ref.\ \cite{KP 99}. The
correspondence principle being used as a guide, a thorough understanding
of the observable features of the classical theory proved to be
basic for the presentation of the quantum theory. However, not all observable
structures of the classical theory identified so far have been
implemented in the quantum theory. One particular such structure
of the classical theory, the laws for the merging of two closed
strings in a single such string and, vice versa, for the splitting
of a single closed string into two closed strings, or rather the
quantization of these laws, is the topic of the present communication.

The laws alluded to expand the Huygens-Newtonian laws of impact for the
collision of two freely moving bodies. In either case, after some
arithmetics, information about the incoming state of the system
produces partial information about the resulting state after the
merging or the splitting of the string(s) and after the collision,
respectively. In contrast to the Huygens-Newtonian laws of impact, the laws
for the string are non-additive in general. Examples of such
classical non-additive composition or decomposition laws for the
closed strings moving in (2+1)--dimensional space-time can be found
in Ref. \cite{KP 95}. There a basic strategy -- also valid for higher
dimensional space-time -- has been described on how to trace and
determine such laws. Along those lines, J. Gro\ss er has specified
the first genuinely non-additive composition laws for the merging
of two closed strings moving in (3+1)--dimensional space-time
Ref.\ \cite{Grosser}.

We shall concentrate on these particular classical
composition laws and show that all (pre-) conditions for their
implementation in the corresponding quantum theory are satisfied,
most important, their adherence to the isomorphic generalized
commutation relations for the string branches involved.

\section*{Classical Theory Part One}

We consider a single closed bosonic massive Nambu--Goto string
which forms an infinitely differentiable space-like closed curve
without double points in (3+1)--dimensional Minkowski space-time
and which is furnished with a smooth bundle of planes tangent to
its trajectory surface, one plane for every point on the curve.
Let its rest mass be denoted by the symbol $\gm$: $\gm>0$. 
\\
As
long as the string doesn't develop double points and doesn't
make contact with other closed strings which might be around,
it sweeps over a two-dimensional smooth tube-shaped branch of
its world-surface. At every point of this branch the corres%
ponding tangent space contains both time-like and space-like
vectors. In order to parametrize the smooth tube-shaped branch of the
string world-surface, we choose a foliation of the Minkowski space by
smooth space-like hypersurfaces (labeled by a forward time-like
parameter $\tau$) intersecting the tube-shaped branch in space-like
smooth closed curves ${\cal C}_{\tau}$ without double points. The closed
curves ${\cal C}_{\tau}$ are the integral curves of a smooth space-like
vector field $\partial _\sigma$ on the tube-shaped branch. The
additional choice of a smooth forward time-like vector field $\partial
_\tau$ on the branch 
and an explicit $\sigma$-parametrization of a single curve  
${\cal C}_{\tau_0}$ complete the construction.

Each space-like leaf of the foliation with label $\tau$ carries a
Riemannian metric induced from the Minkowski space. It also contains
the closed curve ${\cal C}_{\tau}$. Inscribe into ${\cal C}_{\tau}$
within the leaf with label $\tau $ the
minimal surface of least area ( with respect to the Riemannian metric ).
An inward and outward directed normal
vector field on ${\cal C}_{\tau}$ tangent to this minimal surface
can be defined. This implies that an
inside and an outside, hence an orientation of the branch, can be
defined: the parametrization of the branch is said to have positive orientation
if the smooth coordinate vector fields $\pa_\tau$ and
$\pa_\sigma$ (in this order!) at any common point of the branch
follow each other clockwise on the outer surface.

The infinitesimal generators of observable symmetry transformations
of this system form a Poisson algebra. Information about the numerical
values of all elements of this algebra -- provided certain
compatibility requirements are met -- permits to reconstruct the
tube-shaped branch apart from its position in the direction of
the energy-momentum vector of the string. The algebra itself has
the following structure: the Poisson Lie algebra of the infinitesimal
generators of external (rigid) Poincar\'e transformations acts on
the Poisson combinatorial $*$--algebra $\gh$ of the infinitesimal
generators of internal (flexing) symmetry transformations. It is
sufficient to specify the algebra $\gh$ in the momentum rest frame
of the string. The algebra $\gh$ turns out to be the sum of two
Poisson commuting $*$--subalgebras $\gh^+$ and $\gh^-$ which are
isomorphic up to a global sign in the ``structure constants''.
Thus it suffices to present just one of the two subalgebras, say
$\gh^-$, the algebra built from the ``right-movers'', \ie\ the
right-moving combination of the canonical string variables:
$u^-_\mu(\tau,\sigma) := p_\mu(\tau,\sigma) - \frac1{2\pi\alpha'}
\, \pa_\sigma x_\mu(\tau,\sigma)$,
the symbol $\alpha'$ denoting the inverse string tension.
The elements of $\gh^-$ are obtained from the eigenvalues of a
parameter-dependent monodromy matrix corresponding to a system
of linear partial differential equations (containing infinitely
many free parameters). The integrability conditions of this
system are equivalent to the equations of motion for the
right-movers.

The algebra, so obtained, is graded under the Poisson bracket
operation
$\{ \cdot , \cdot \} := 2\pi\alpha'$
times the canonical Poisson bracket operation:
\[  \gh^- = \bigoplus_{l=0}^\infty \; \GV^l ,\qquad
    \{ \GV^{l_1} , \GV^{l_2} \} \subset \GV^{l_1 + l_2} ,\qquad
    \GV^{l_1} \cdot \GV^{l_2}   \subset \GV^{l_1 + l_2 + 1} ,\qquad
    l_1, l_2 \in \N_0  .\]
Moreover, the finite dimensional vector spaces $\GV^l$ are invariant
under the parity transformation. Hence, they are decomposable into
parity eigenspaces
\[  \GV^l = \GV^l_+ \oplus \GV^l_-  .\]
The vector space $\GV^0$ consists of parity even elements only:
$\GV^0 \equiv \GV^0_+$. It is three-dimensional and forms a
subalgebra isomorphic to the Lie algebra $so(3)$ with generators
$J_{1,-1}$, $J_{1,0}$ and $J_{1,+1}$ in an angular momentum basis,
or rather a spin basis. Each one of the vector spaces $\GV^l_\pm$
is invariant under the Poisson action of $\GV^0$. Hence each
$\GV^l_\pm$ carries a linear representation of $so(3)$ and can be
decomposed into a direct sum of isotypical components corresponding
to the spin $j\in\N_0$
\[  \GV^l_+ = \bigoplus_{j=0}^{l+1} \: \GV^l_{j,+} ,\qquad\qquad
    \GV^l_- = \bigoplus_{j=0}^{l+1} \: \GV^l_{j,-} .\]
Some of the isotypical components may be trivial.

In the following we shall only be concerned with the vector spaces
$\GV^l$ of degree $0 \le l \le 2$. Their decompositions can be taken
from the following list:
\[  \GV^0 = \GV^0_{1,+} \]
with $\dim\,\GV^0_{1,+} = 3$;
\[  \GV^1 = \GV^1_+ \oplus \GV^1_- , \qquad
    \GV^1_+ = \GV^1_{0,+} \oplus \GV^1_{2,+} \qquad
    \GV^1_- = \GV^1_{1,-} \oplus \GV^1_{2,-}  \]
with respective dimensions: $\dim \GV^1_+ = 12 = 2 \oplus 10$ and
$\dim \GV^1_- = 8 = 3 \oplus 5$;
\[  \GV^2 = \GV^2_+ \oplus \GV^2_- , \qquad
    \GV^2_+ = \GV^2_{1,+} \oplus \GV^2_{2,+} \oplus \GV^2_{3,+} \qquad
    \GV^2_- = \GV^2_{0,-} \oplus \GV^2_{1,-} \oplus \GV^2_{2,-}
                          \oplus \GV^2_{3,-} \]
with respective dimensions: $\dim \GV^2_+ = 49 = 18 \oplus 10 \oplus
21$ and $\dim \GV^2_- = 43 = 2 \oplus 12 \oplus 15 \oplus 14$.

We organize all elements of the algebra $\gh^-$ in the shape of
irreducible tensor variables
$\mathscr{O}_j = \{ \, \mathscr{O}_{j,m} \,|\, -j \le m \le +j \,\}$
(under the action of $O(3)$), \eg\
\[  J_1 = \{\, J_{1,m} \,|\, -1 \le m \le +1 \,\} \]
\[  ( J_1 \cdot J_1 )_0 = ( J_1^2 )_0 = \sum_{m_1} \sum_{m_2} \:
    \langle 1,m_1 ; 1,m_2 | 0,0 \rangle \; J_{1,m_1} J_{1,m_2} \]
\[  ( J_1 \cdot J_1 )_2 = ( J_1^2 )_2 = \{\, ( J_1^2 )_{2,m} =
    \sum_{m_1} \sum_{m_2} \: \langle 1,m_1 ; 1,m_2 | 2,m \rangle \;
    J_{1,m_1} J_{1,m_2} \:|\, -2 \le m \le +2 \,\} , \]
the symbols $\langle j_1,m_1 ; j_2,m_2 | j,m \rangle$ denoting the
Clebsch--Gordan coefficients with the conventions of Condon and
Shortley Ref.\ \cite{Condon}.

Together with some other generators of the algebra $\gh^-$ besides
$J_1$, \ie\ some other irreducible tensor variables generating the
algebra $\gh^-$ via multiplication and Poisson bracket operation,
to wit the parity even generators $B_0^{(1)}$ and $T_2$ along with
the parity odd generators $S_1$ and $S_2$, a basis of the above listed
vector spaces is given in terms of the components of irreducible
tensor variables by
\begin{align*}
\GV^0_{1,+} &: \quad J_1  \\
\GV^1_{0,+} &: \quad B_0^{(1)}, \, ( J_1^2 )_0  \qquad\qquad
\GV^1_{2,+}  : \quad T_2 ,\, ( J_1^2 )_2 \\
\GV^1_{1,-} &: \quad S_1  \qquad\qquad\qquad\quad\;\;\,
\GV^1_{2,-}  : \quad S_2  \\
\GV^2_{1,+} &: \quad \{ T_2,T_2 \}_1 ,\, \{ S_2,S_1 \}_1 ,\,
                     \{ S_1,S_1 \}_1 ,\, ( J_1 \cdot B_0^{(1)} )_1 ,\,
                     ( J_1 \cdot T_2 )_1 ,\, ( J_1 \cdot ( J_1^2 )_0 )_1 \\
\GV^2_{2,+} &: \quad \{ S_2,S_1 \}_2 ,\, ( J_1 \cdot T_2 )_2 \\
\GV^2_{3,+} &: \quad \{ S_2,S_1 \}_3 ,\, ( J_1 \cdot T_2 )_3 ,\,
                     ( J_1 \cdot ( J_1^2 )_2 )_3
\end{align*}
and, finally,
\begin{align*}
\GV^2_{0,-} &: \quad \{ T_2,S_2 \}_0 ,\, ( J_1 \cdot S_1 )_0 \\
\GV^2_{1,-} &: \quad \{ T_2,S_2 \}_1 ,\, \{ T_2,S_1 \}_1 ,\,
                     ( J_1 \cdot S_2 )_1 ,\, ( J_1 \cdot S_1 )_1 \\
\GV^2_{2,-} &: \quad \{ T_2,S_1 \}_2  ,\, ( J_1 \cdot S_2 )_2 ,\,
                     ( J_1 \cdot S_1 )_2  \\
\GV^2_{3,-} &: \quad \{ T_2,S_1 \}_3  ,\, ( J_1 \cdot S_2 )_3 .
\end{align*}
Here the symbols $( \SO_{j_1} \cdot \SO_{j_2} )_j$ and $\{ \SO_{j_1} ,
\SO_{j_2} \}_j$ stand for the irreducible tensor variables
$\{\, ( \SO_{j_1} \cdot \SO_{j_2} )_{j,m} \,|\, -j \le m \le +j \,\}$
and
$\{\, \{ \SO_{j_1} , \SO_{j_2} \}_{j,m} \,|\, -j \le m \le +j \,\}$,
with
\[  ( \SO_{j_1} \cdot \SO_{j_2} )_{j,m} =
    \sum_{m_1} \sum_{m_2} \: \langle j_1,m_1 ; j_2,m_2 | j,m \rangle \;
    \SO_{j_1,m_1} \cdot \SO_{j_2,m_2}  \]
and
\[  \{ \SO_{j_1} , \SO_{j_2} \}_{j,m} =
    \sum_{m_1} \sum_{m_2} \: \langle j_1,m_1 ; j_2,m_2 | j,m \rangle \;
    \{ \SO_{j_1,m_1},\SO_{j_2,m_2} \} , \]
respectively.

The remaining non-trivial Poisson brackets between the generators
$B_0^{(1)}$, $S_1$, $S_2$ and $T_2$, \ie\ the ones not listed as
basis elements of $\GV^2_+$ and $\GV^2_-$, are given by
\[   \begin{array}{lcl}
\{T_2,T_2\}_{3}  &=&  - i\,\{S_2,S_1\}_{3} +
                       i\,16\,( J_1 \cdot ( J_1^2 )_2 )_3  \,, \\
\{S_2,S_2\}_{3}  &=&    i\,2\,\{S_2,S_1\}_{3} - 
                       i\,8\,(J_1\cdot T_2)_{3} -
                       i\,48\,( J_1 \cdot ( J_1^2 )_2 )_3 \,, \\
\{S_2,S_2\}_{1}  &=&  - i\,{\sqrt{{\frac{2}{3}}}}\;\{S_2,S_1\}_{1} -
                          {\frac{1}{6}}\,{\sqrt{5}}\;\{S_1,S_1\}_{1} -
                       i\,16\,{\sqrt{{\frac{2}{3}}}}\;(J_1\cdot T_2)_{1} -\\
                 & &   i\,32\,{\sqrt{{\frac{2}{15}}}}\;
                              (J_1\cdot (J_1^{2})_{0})_{1}\,, \\
\{ B_0^{(1)},T_2 \}_2 &=& i\,{\sqrt{6}}\; \{ S_2,S_1 \}_2 \,; \\
\{T_2,S_2\}_{4}  &=&  0 \,, \\
\{T_2,S_2\}_{3}  &=&  i\,\{T_2,S_1\}_{3} + i\,8\,(J_1\cdot S_2)_{3}\,, \\
\{T_2,S_2\}_{2}  &=&  -{\frac{i}{3}}\,{\sqrt{{\frac{7}{2}}}}\;
                       \{T_2,S_1\}_{2} -
                       i\,{\frac{4}{3}}\,{\sqrt{14}}\;(J_1\cdot S_2)_{2}\,, \\
\{B_{0}^{(1)},S_{2} \}_{2} &=& - i\,2\,{\sqrt{{\frac{2}{3}}}}\;
                       \{ T_{2},S_{1} \}_{2} -
                       i\,4\,{\sqrt{{\frac{2}{3}}}}\;
                       (J_{1}\cdot S_{2})_{2} +
                       12\,(J_{1}\cdot S_{1})_{2} \,, \\
\{B_{0}^{(1)},S_{1} \}_{1} &=& - i\,6\,{\sqrt{{\frac{2}{5}}}}\;
                       \{ T_{2},S_{2} \}_{1} +
                       2\,{\sqrt{{\frac{3}{5}}}}\;\{T_{2},S_{1}\}_{1} -
                       24\,{\sqrt{{\frac{3}{5}}}}\;(J_{1}\cdot S_{2})_{1} +\\
                 & &   i\,12\,{\sqrt{2}}\;(J_{1}\cdot S_{1})_{1} \, .
\end{array}   \]
In the sequel these relations will be referred to as generalized
Poisson commutation relations.

\section*{Classical Theory Part Two}

Next, we consider two massive closed strings (of the type described
before and of common positive orientation) making contact in a distinguished
space-time point $x^{(0)}$ and subsequently forming a single
third closed string of the same type. That is, the two tube-shaped
branches (numbered $[1]$ and $[2]$) merge in the space-time point
$x^{(0)}$ into a single third such branch (numbered $[0]$). The
triple of branches constitutes a ``string vertex''. According to
the laws of impact, the energy momentum vectors of the two merging
strings $P_\mu^{[1]}$ and $P_\mu^{[2]}$, respectively, when added
yield the energy momentum vector $P_\mu^{[0]}$ of the merged string.
The same goes for the infinitesimal generators of the restricted
Lorentz transformations $M_{\mu\nu}$:
$M_{\mu\nu}^{[1]} + M_{\mu\nu}^{[2]} = M_{\mu\nu}^{[0]}$.
This formulation refers to a fixed Lorentzian coordinate system in the
(3+1)--dimensional Minkowski space. Since the quantities
$P_\mu^{[\cdot]}$ and $M_{\mu\nu}^{[\cdot]}$ do not vary
while the pertinent strings sweep over their branches, these
quantities can just as well be assigned to the pertinent
branches rather than to the strings. The same goes for the
infinitesimal generators of the internal symmetry transformations:
$\JE{\cdot}$, $\BNE{\cdot}$, $\TZ{\cdot}$, $\SZ{\cdot}$, $\SE{\cdot}$,
$\dots$. They, too, are piece-wise conserved quantities.

The infinitesimal generators of rotations in the momentum rest
frame of the branch $k$, $k=0,1,2$, are given by $\frac1{i\gm^{[k]}}$
times the spatial components of the corresponding Pauli--Lubanski
vector
\[  \SW^{[k]\mu} = {\textstyle \frac{i}2} \, \epsilon^{\mu\rho\sigma\nu} \,
    M_{\rho\sigma}^{[k]} \, P_\nu^{[k]} ,\qquad\qquad\qquad
    \epsilon^{0123} = + 1  \]
in the rest frame of the branch $k$: $\SW^{[k]j}_{\quad | k}$,
$j=1,2,3$ ($\SW^{[k]0}_{\quad | k} = 0$; the bar followed by a
branch index $k$ indicates that the respective quantity is to be
evaluated in the rest frame of the branch [k]).

From the law of addition of the infinitesimal generators
$M_{\mu\nu}^{[\cdot]}$, all that is left as a law for the
composition of the spin momenta (\ie\ the angular momenta
corresponding to rotations in the various rest frames) is
\[  \SW^{[0]\mu} \, P_\mu^{[1]} =
    ( \SW^{[2]\mu} - \SW^{[1]\mu} ) \, P_\mu^{[0]} =
    \SW^{[2]\mu} \, P_\mu^{[1]} - \SW^{[1]\mu} \, P_\mu^{[2]} .\]
Analogously, there is only one (additive) composition law
relating the basis elements $\JE{k}$ of the three-dimensional
subspaces $\GV^{[k]0} = \GV^{[k]0}_+$ of $\gh^{[k]-}$, $k=0,1
,2$, $\JE{k} = \{\, J^{[k]}_{1,m} \,|\, -1 \le m \le +1 \,\}$
with $J^{[k]}_{1,m}$ acting like the spin momenta in $\gh^{[k]-}$,
to wit:
\[  ( \AE \cdot \PE{1}{0} )_0 = 0 .\]
Here $\AE$ stands for the following irreducible $j=1$ tensor
variable in the rest frame of the branch 0
\[  \AE := - i \, \,\JE{0} \, \potimes +
            \WE{1}{0} \, \poz + \WE{2}{0} \, \poe \]
with ($k=1,2$; $[\neq 1] := [2]$, $[\neq 2] := [1]$)
\[  \WE{k}{0} := i \, \big[ \JE{k} \, \gm^{[k]} +
    ( \JE{k} \cdot \PE{\neq k}{k} )_0 \,
    \delta^{[k]} \, \PE{k}{0}   \big]
    , \qquad\qquad  
    \delta^{[k]} := \frac{\gm^{[k]}}{\mn}
                    \frac{\sqrt{3}}{P_{0|0}^{[k]} + \gm^{[k]} } \, .\]
The laws of impact and the above composition law are examples of
``selection rules'' for the string vertex. There are infinitely
many more selection rules relating basis elements of subspaces of
$\gh^{[k]-}$ (and $\gh^{[k]+}$) of higher degrees to each other,
$k=0,1,2$. However, in general apart from the elements of the
Poincar\'e algebra and the isomorphic Poisson algebras $\gh^{[k]-}$,
$k=0,1,2$, and $\gh^{[k]+}$, $k=0,1,2$, respectively, they involve
a variable which encodes the phase of ``breathing'' of string 1
relative to the phase of ``breathing'' of string 2 just before
the two strings merge in the point $x^{(0)}$ to form the string 0.

For simplicity, we focus on selection rules which do not involve
that kind of variable, which do not involve elements of the
algebras $\gh^{[k]+}$, $k=0,1,2$, and which relate basis elements
of the subspaces $\GV^{[k]1}$, $k=0,1,2$, and $\GV^{[k]2}$, $k=0,1,2$,
respectively. Under these restrictions our analysis (first classical,
then quantal) will be exhaustive.

In his diploma thesis Ref.\ \cite{Grosser} J.\ Gro\ss er has identified the
only two ``new'' (tensorial) composition laws (of the restricted type)
relating basis elements of the subspaces $\GV^{[k]1}$, $k=0,1,2$,
to each other. The set of these two laws is 
symmetric under the following substitution  which we call 
``{\em formal} crossing'': 
replace the real process describing the merging of branch [1] with branch [2]
to form branch [0] by the non-real process describing the
merging of branch [1] with crossed branch [0] to form crossed branch
[2]. Here, the crossed branch is obtained from the original branch by
replacing
$p_\mu(\tau,\sigma)$ by $-p_\mu(\tau,\sigma)$ and by
subsequently inverting the orientation of the corresponding curves
${\cal C}_\tau^{[\cdot]}$.

Gro\ss er has presented these laws in a manifestly Lorentz covariant
form. For our purposes, however, it is advantageous to evaluate them
in the rest frame of the merged branch $[0]$ and to state the
dependence of the contributions from the branches $[1]$ and $[2]$ on
the elements of their respective rest frame algebras $\gh^{[1]-}$
and $\gh^{[2]-}$ explicitly. For the time being, we regard the
composition laws as numerical relations which follow from continuity
requirements imposed on the aforementioned monodromy matrices and
from factorization properties of the latter ones.

We introduce the following notation ($k=1,2$):
\begin{equation*}
\begin{array}{lcl}
\XZC{0}  &=&  - \imag \,{\sqrt{6}}\;(\SZ{0}\cdot\PE{1}{0})_{2} +
              (\SE{0}\cdot\PE{1}{0})_{2}  \,, \\[2ex]
\XZC{k}  &=&  - \imag \,{\sqrt{6}}\;(\SZ{k}\cdot\PE{\neq k}{k})_{2} +
              (\SE{k}\cdot\PE{\neq k}{k})_{2} \,, \\[2ex]
\ZZC{0}  &=&  - 2\,{\sqrt{3}}\;(\TZ{0}\cdot\PN{1}{0})_{2} +
                2\,{\sqrt{21}}\;(\TZ{0}\cdot\PM{2}{1}{0})_{2} -
                (\BNE{0}\cdot\PM{2}{1}{0})_{2} \,, \\[2ex]
\ZZC{k}  &=&  - 2\,{\sqrt{3}}\;(\TZ{k}\cdot\PN{\neq k}{k})_{2} +
                2\,{\sqrt{21}}\;(\TZ{k}\cdot\PM{2}{\neq k}{k})_{2} -
                (\BNE{k}\cdot\PM{2}{\neq k}{k})_{2} \,, \\[2ex]
\Xi^{[k]}_{2|0}     &=&
    - \frac13 \, \ZZC{k}
    - \frac13 \, \XZC{k} \, \frac{\mn}{\gm^{[k]}} \,
      ( P_{0|0}^{[k]} + 2 \gm^{[k]} ) - \\[1ex]
&&
      \frac13 \, \delta^{[k]} \: \Big[
      \sqrt{7} \; ( \XZC{k} \cdot \PM{2}{\neq k}{k} )_2
      - 2 \, \sqrt{5} \; ( \XZC{k} \cdot \PM{2}{\neq k}{k} )_0 \:
        \frac{\PM{2}{\neq k}{k}}{\PN{\neq k}{k}}  \: \Big]  - \\[1ex]
&&
      \frac43 \, \JEQZ{k} \, \Big( \frac{\mn}{\gm^{[k]}} \Big)^2 \,
      \frac{[ \PO{k} + \gm^{[k]} ]^3 [ \PO{k} - \gm^{[k]} ]}{(\PO{k})^2} -\\[1ex]
&&
      \frac4{\sqrt{3}} \; \Big[
      \sqrt{7} \; ( \JEQZ{k} \cdot \PM{2}{\neq k}{k} )_2 + 
      2 \, \JEQN{k} \, \PM{2}{\neq k}{k}  \Big] \,
      \frac{\big[ (\PO{k})^2 - \gm^{[k]} \PO{k} + (\gm^{[k]})^2 \big]}
           {(\PO{k})^2} + \\[1ex]
&&
      4 \, \sqrt{3} \; \big[ ( \JE{k} \cdot \PE{\neq k}{k} )_0 \big]^2 \,
      \frac {\PM{2}{\neq k}{k}}{\PN{\neq k}{k}}
      \frac{[ \PO{k} - \gm^{[k]} ]^2}{(\PO{k})^2} \,, \\[2ex]
\Theta^{[k]}_{2|0}  &=&
      \ZZC{k} \, \PO{k} +
      \sqrt{3} \, \Big[
      - \XZC{k} \, \PN{\neq k}{k}
      + \sqrt{7} \; ( \XZC{k} \cdot \PM{2}{\neq k}{k} )_2 - \\[1ex]
&&
      2 \, \sqrt{5} \; ( \XZC{k} \cdot \PM{2}{\neq k}{k} )_0 \,
      \frac{\PM{2}{\neq k}{k}}{\PN{\neq k}{k}}  \Big] \,
      \frac{\gm^{[k]}}{\mn} +\\[1ex]
&&
      4 \, \sqrt{3} \; \Big[
      - \JEQZ{k} \, \PN{\neq k}{k} +
      \sqrt{7} \; ( \JEQZ{k} \cdot \PM{2}{\neq k}{k} )_2 +
      2 \, \JEQN{k} \PM{2}{\neq k}{k}  \Big] \times \\[1ex]
&&
      \frac{[ (\PO{k})^2 - (\gm^{[k]})^2 ]}{\PO{k}} +
      36 \, \big[ ( \JE{k} \cdot \PE{\neq k}{k} )_0 \big]^2 \,
      \PM{2}{\neq k}{k} \, \frac1{\PO{k}} \, 
      \Big( \frac{\gm^{[k]}}{\mn} \Big)^2 \,, \\[2ex]
\BE   &=&   - i \, \JE{0} \, \pominus + \WE{1}{0} - \WE{2}{0} .
\end{array}
\end{equation*}
{\it Note}: The irreducible tensor variables $\Xi^{[k]}_{2|0}$ and
$\Theta^{[k]}_{2|0}$ are linear in $\XZC{k}$ and $\ZZC{k}$, they
are quadratic in $\JE{k}$, and they are mass and momentum dependent.

With these abbreviations the composition laws read
\begin{align}
\label{comp1}
\XZC{0} &= \Big[ \Xi^{[1]}_{2|0} - \Xi^{[2]}_{2|0} \Big] \, 
{\textstyle \frac1{\mn}} -
           4 \, (\AE^2)_2 \, {\textstyle \frac{\pominus}{\potimesq}} +
           8\, ( \AE \cdot \BE )_2 \, {\textstyle \frac1{\potimes}} \,, \\
\label{comp2}
\ZZC{0} &= \Big[ \Theta^{[1]}_{2|0} + \Theta^{[2]}_{2|0} \Big] \, 
{\textstyle \frac1{\mn}}
           + 12 \, (\AE^2)_2 \, {\textstyle \frac{1}{\potimes}} \,.
\end{align}
The left and the right hand sides of the first law are antisymmetric
under the interchange of the branches $[1]$ and $[2]$, the left and
the right hand sides of the second law are symmetric (note: $\PE{2}{0}
= - \PE{1}{0}$).

There are no ``kinematical'' constraints on the irreducible tensor
variable $\XZC{k}$, $k=0,1,2$. Hence, the components of $\XZC{0}$
furnish 5 independent linear combinations of the components of the
irreducible tensor variables $\SZ{0} \in \GV^{[0]1}_-$ and $\SE{0}
\in \GV^{[0]1}_-$ which -- with the help of $\JE{0} \in \GV^{[0]0}_+$
-- can be expressed in terms of the basis elements of the subspaces
$\GV^{[1]l}$ and $\GV^{[2]l}$, $l=0,1$.

In contrast, there do exist ``kinematical'' constraints on the
irreducible tensor variable $\ZZC{k}$, $k=0,1,2$, to wit:
\[  ( \ZZC{0} \cdot \PM{2}{1}{0} )_1 = 0 ,\qquad\qquad
    ( \ZZC{k} \cdot \PM{2}{\neq k}{k} )_1 = 0 ,\qquad
    k=1,2 .\]
They imply, that there are only 3 independent linear combinations
of the components of the irreducible tensor variables $\BNE{0} \in
\GV^{[0]1}_+$ and $\TZ{0} \in \GV^{[0]1}_+$ which -- with the help
of $\JE{0} \in \GV^{[0]0}_+$ -- can be expressed in terms of the
basis elements of the subspaces $\GV^{[1]l}$ and $\GV^{[2]l}$, $l=0,1$.

We express the basis elements $\SE{0}$ and $\SZ{0}$ in terms of
$\XZC{0}$ and $\se := 2 \, ( \SZ{0} \cdot \PE{1}{0} )_1$,
likewise the basis elements $\BNE{0}$ and $\TZ{0}$ in terms of
$\ZZC{0}$ and $\te := 2 \, ( \TZ{0} \cdot \PE{1}{0} )_1$:
\begin{equation*}
    \begin{array}{lcl}
        \SE{0}  &=&  i \, \frac{1}{3}\,{\sqrt{10}}\;
             (\se\cdot\frac{\PE{1}{0}}{\PN{1}{0}})_{1} +
             \frac{3}{\sqrt{5}} \;
             (\XZC{0}\cdot\frac{\PE{1}{0}}{\PN{1}{0}})_{1} -
             \frac{1}{6} \, \sqrt{\frac{7}{3}}\;
             (\XZC{0}\cdot\frac{\PM{3}{1}{0}}{{\PN{1}{0}}^2})_{1} \,, \\
\SZ{0}  &=&  \frac{3}{2\,{\sqrt{5}}} \;
             (\se\cdot\frac{\PE{1}{0}}{\PN{1}{0}})_{2} -
             \frac{1}{12} \, {\sqrt{\frac{7}{3}}} \;
             (\se\cdot\frac{\PM{3}{1}{0}}{{\PN{1}{0}}^2})_{2} -
             i \, \frac{1}{5}\,{\sqrt{2}}\;
             (\XZC{0}\cdot\frac{\PE{1}{0}}{\PN{1}{0}})_{2} + \\
        & &  i \, \frac{1}{6} \, {\sqrt{\frac{7}{5}}} \;
             (\XZC{0}\cdot\frac{\PM{3}{1}{0}}{{\PN{1}{0}}^2})_{2} \,, \\
\BNE{0} &=&  \frac{{\sqrt{15}}}{2} \;
             (\te\cdot\frac{\PE{1}{0}}{\PN{1}{0}})_{0} -
             \frac{{\sqrt{5}}}{2} \;
             (\ZZC{0}\cdot\frac{\PM{2}{1}{0}}{{\PN{1}{0}}^2})_{0} \,, \\
\TZ{0}  &=&  \frac{3}{2\,{\sqrt{5}}} \;
             (\te\cdot\frac{\PE{1}{0}}{\PN{1}{0}})_{2} -
             \frac{1}{12} \, {\sqrt{\frac{7}{3}}} \;
             (\te\cdot\frac{\PM{3}{1}{0}}{{\PN{1}{0}}^2})_{2} -
             \frac{1}{12\,{\sqrt{3}}} \;
             (\ZZC{0}\cdot\frac{1}{\PN{1}{0}})_{2} + \\
        & &  \frac{1}{24} \, {\sqrt{\frac{7}{3}}}\;
             (\ZZC{0}\cdot\frac{\PM{2}{1}{0}}{{\PN{1}{0}}^2})_{2} \, .
         \end{array}
\end{equation*}
There do not exist relations (of the restricted type) of the
tensor variables $\se$ and $\te$ to the elements of the
subspaces $\GV^{[k]1}$, $k=1,2$.

We have searched for ``new'' relations among the basis elements of the
subspaces $\GV^{[k]2}$, $k=0,1,2$, relations of the restricted
type, which are not merely the result of multiplication of
relations already known. We found out that there are exactly
21 independent components of such composition laws. In more
detail, our findings are: With the help of $\JE{0}$, 11 (10)
independent linear combinations of the basis elements of
$\GV^{[0]2}_+$ ($\GV^{[0]2}_-$) can be expressed in terms of
the basis elements of the subspaces $\GV^{[k]l}$, $k=1,2$;
$\;0 \le l \le 2$.

In order to present these composition laws, one need not start
from scratch. They can be obtained from the previous ones (\cf\
equations (\ref{comp1}) and (\ref{comp2})) in a systematic way by
Poisson bracket operations. Also, they can be analyzed by this
tool.

This leads us to the second part of this section dedicated to
the (Poisson) algebraic aspects of the relations under discussion.

Stated as above in terms of irreducible tensor multiplets, the
covariance of the left hand sides and the right hand sides,
respectively, of the composition laws (\ref{comp1}) and (\ref{comp2})
under rotations in the momentum rest frame of the merged branch is
obvious. The rotations are generated by $\frac1{i \mn}$ times the
spatial components $\SW_{j|0}^{[0]}$ of the Pauli--Lubanski vector
in the aforementioned rest frame.

As far as the elements of the Poisson algebra $\gh^{[0]-}$ are
concerned, the very same rotations are also generated by the
(Cartesian) components of $\JE{0}$. A priori, the Poisson action
of $\JE{0}$ on the components of $\PE{1}{0} = - \PE{2}{0}$ and
on the elements of $\gh^{[k]-}$, $k=1,2$, is not defined. We are
free to define the Poisson action of $\JE{0}$ on these objects
provided the definition complies with the formal properties of
Poisson algebras and provided it ensures that the action of
$\JE{0}$ takes valid relations into valid relations. In particular,
this implies that the action of $\JE{0}$ on all irreducible tensor
variables in the rest frame of the merged branch such as $\PE{1}{0}
= - \PE{2}{0}$, $\WE{k}{0}$, $k=1,2$, $\Xi_{2|0}^{[k]}$, $\Theta_
{2|0}^{[k]}$, $k=1,2$, is identical with the action of
$  \frac1{i \mn} \, \SW_{1|0}^{[0]} =
    \frac1{i\mn} \, ( \SW_{1,-1}^{[0]}, \SW_{1,0}^{[0]}, \SW_{1,+1}^{[0]} ) =
    \frac1{i\mn} \, ( \frac1{\sqrt{2}} \, ( \SW_{1|0}^{[0]} -
                                      i \, \SW_{2|0}^{[0]} ),
                                           \SW_{3|0}^{[0]},
                  \frac{-1}{\sqrt{2}} \, ( \SW_{1|0}^{[0]} +
                                      i \, \SW_{2|0}^{[0]} ))$.
Since the elements of $\gh^{[k]-}$, $k=0,1,2$, apart from momentum
densities, encode coordinate {\it differences} only, the components
of the generator $P_\mu^{[0]}$ of joint rigid translations of all
three branches have vanishing Poisson brackets with each and every
one of these elements. Thus, $\pon = \mn$ plays the r\^ole of a
common central element. The components of the vectors $P_\mu^{[1]}$
and $P_\mu^{[2]}$ not only Poisson commute among each other, but
they have also vanishing Poisson brackets with all elements of
$\gh^{[k]-}$, $k=1,2$. The same goes for their projections $\PE{1}{0}
= - \PE{2}{0}$, $\poe$ and $\poz$. Minkowski space-time causality
implies that the elements of $\gh^{[1]-}$ and the elements of
$\gh^{[2]-}$ mutually Poisson commute. The Poisson brackets of
$\PE{1}{0} = - \PE{2}{0}$ and $\poe - \poz$ ($\poe + \poz = \mn$ !)
with the generators $\BNE{0}$, $\SE{0}$, $\SZ{0}$, $\TZ{0}$, $\dots$
of the algebra $\gh^{[0]-}$ are not defined a priori. In fact, their
definition is a rather delicate matter. We get a first hint as to
their definition from the composition laws (\ref{comp1}) and
(\ref{comp2}) focussing on their algebraic rather than their
numerical aspects. Defining the Poisson brackets $\{ \XZC{0},
\PE{1}{0} \}_L$, $\{ \ZZC{0},\PE{1}{0} \}_L$, $L=1,2,3$,
$\{ \XZC{0},\pominus \}_2$ and $\{ \ZZC{0},\pominus \}_2$ in
accordance with the right hand sides of the composition laws,
we are led to the following specifications
\[  \{ \XZC{0},\PE{1}{0} \}_L = \begin{cases}
    4 \,{\sqrt{\frac{10}{3}}}\;(\BE\cdot\PE{1}{0})_{1} \qquad\qquad &
    \text{for $L=1$,} \\
    4 \,{\sqrt{6}}\;(\BE\cdot\PE{1}{0})_{2} &
    \text{for $L=2$,} \\
    0 & \text{for $L=3$,} \end{cases}   \]

\[  \{ \XZC{0},\pominus \}_2 = 0 ; \]
\[  \{ \ZZC{0},\PE{1}{0} \}_L = \begin{cases}
    4 \, {\sqrt{30}} \; (\AE\cdot\PE{1}{0})_{1} \qquad\qquad  &
    \text{for $L=1$,} \\
    12 \, {\sqrt{6}} \; (\AE\cdot\PE{1}{0})_{2}  &
    \text{for $L=2$,} \\
    0 & \text{for $L=3$,} \end{cases}   \]
\[  \{ \ZZC{0},\pominus \}_2 = 0 . \]
Next we shall test the consistency of the above specifications with
the generalized Poisson commutation relations for the three branches,
the three corresponding sets of relations being isomorphic to each
other. As we already found out, there
are exactly 10 (11) ``new'' composition laws in the $P$--linear
span of the subspaces $\GV^{[k]2}$, $k=0,1,2$, antisymmetric
(symmetric) under the exchange of the branches $[1]$ and $[2]$
(for the concept of $P$--linearity and related concepts consult
Ref.\ \cite{PR 86}).
The Poisson bracket operation based on the assignments given so
far must take valid relations into valid relations.

Naively, \ie\ not paying attention to the generalized Poisson commutation
relations, the 5 (3) independent components of the composition laws
$\XZC{0} = \dots$ ($\ZZC{0} = \dots$) by Poisson bracket operations
should produce 15 (13) valid ``new'' composition laws of the antisymmetric
and symmetric type, respectively. Hence, the components of the
composition laws so produced are subject to at least 5 (2) constraints
due to the generalized Poisson commutation relations.

For reasons of manifest $O(3)$--invariance,
we do {\it not} solve the kinematical constraint on the composition
law (\ref{comp2}) for the independent corresponding components of
the left hand side $\ZZC{0}$ and the right hand side, respectively.
%
%
%
Instead, we list the $P$--linearly independent combinations of the
Poisson brackets $\{ \ZZC{0} , \XZC{0} \}_L$; $\{ \XZC{0} ,
\XZC{0} \}_L$ and  $\{ \ZZC{0} , \ZZC{0} \}_L$
which are not reduced to $P$--linear combinations of the products
$(\ZZC{0} \cdot \BE )_j$, $(\XZC{0} \cdot \AE )_j$, $(\ZZC{0}
\cdot \AE )_j$ and $(\XZC{0} \cdot \BE )_j$
by the kinematical constraints. We begin with the 15 combinations
involving the Poisson brackets which are antisymmetric under the
interchange of the merging branches. They are identified as
the components of $\{ \ZZC{0} , \XZC{0} \}_L$, $L=0,2,3$ and
the two $P$--linearly independent combinations of the components
of the irreducible tensor $\{ \ZZC{0} , \XZC{0} \}_4$ which are
not contained in the $P$--linear span of the components of the
irreducible tensor $( \{ \ZZC{0} , \XZC{0} \}_4 \cdot \PE{1}{0}
)_3$.
\\
Now we turn to the 13 combinations involving the Poisson brackets
which are symmetric under the interchange of the merging branches.
They are identified as
the components of $\{ \XZC{0} , \XZC{0} \}_L$, $L=1,3$,
the ``projection'' 
$( \{ \ZZC{0} , \ZZC{0} \}_3 \cdot \PM{3}{1}{0} )_0$, and
the two $P$--linearly independent combinations of the components
of the irreducible tensor $\Z_2 := ( \{ \ZZC{0} , \ZZC{0} \}_3 \cdot
\PM{2}{1}{0} )_2$ which are not contained in the $P$--linear span
of the components of the irreducible tensor $( \Z_2 \cdot \PE{1}{0}
)_1$.

Apart from the formal properties of the Poisson brackets, \ie\
antisymmetry, Jacobi identity and Leibniz rule, the only concretely
specified mixed Poisson brackets used for the transport of the
kinematical constraints on the components of $\ZZC{0}$ to those on the
Poisson brackets of the components of $\XZC{0}$ and $\ZZC{0}$
are the ones given above, {\it viz}.\ 
$\{ \XZC{0},\PE{1}{0} \}_L$, $\{ \ZZC{0},\PE{1}{0} \}_L$ and
$\{ \JE{0}, \SO^{[k]}_j \}_L$, $k=1,2$.
These concrete specifications were designed to be compatible with the
composition laws $\XZC{0}=\dots$, $\ZZC{0}=\dots$. Thus the
kinematical constraints on the components of $\{ \ZZC{0} , \XZC{0} \}_L$
and $\{ \ZZC{0} , \ZZC{0} \}_L$ are automatically satisfied if
for $\XZC{0}$ and $\ZZC{0}$ the right hand sides of their composition
laws are inserted.

\vspace{5mm}
Next we exhibit the additional ``dynamical'' constraints (due to the
generalized Poisson commutation relations for the merged branch) on
the 15 + 13 $P$--linear combinations of the Poisson brackets $\{ \ZZC{0}
, \XZC{0} \}_L$; $\{ \XZC{0} , \XZC{0} \}_L$ and $\{ \ZZC{0} , \ZZC{0}
\}_L$ listed above.
\begin{equation*}
\begin{array}{lcl} 
    0  &=&
    i \, (\{\ZZC{0},\XZC{0}\}_{2} \cdot \PM{2}{1}{0})_{0} +
    i \, \frac{10}{7\,{\sqrt{7}}} \,
    (\{\ZZC{0},\XZC{0}\}_{0} \cdot \PN{1}{0})_{0} + \\[1ex]
&&
    \frac{192}{7} \, {\sqrt{\frac{6}{7}}} \;
    ((\XZC{0} \cdot \JE{0})_{2} \cdot \PN{1}{0}\,\PM{2}{1}{0})_{0} +
    i \, \frac{192}{7} \, {\sqrt{\frac{2}{7}}}\;
    ((\XZC{0} \cdot \AE)_{2} \cdot \PM{2}{1}{0})_{0}   \,,   
\end{array} 
\end{equation*} 
\begin{equation*}
\begin{array}{lcl}
    0  &=&
    i \, (\{\ZZC{0},\XZC{0}\}_{3}\cdot\PN{1}{0}\,\PM{2}{1}{0})_{2} -
    i \, {\sqrt{\frac{7}{10}}} \;
    (\{\ZZC{0},\XZC{0}\}_{3}\cdot\PM{4}{1}{0})_{2} + \\[1ex]
&&
    16 \, {\sqrt{3}} \;
    ((\XZC{0}\cdot\JE{0})_{3}\cdot{\PN{1}{0}}^2\,\PM{2}{1}{0})_{2} -
    8 \, {\sqrt{\frac{42}{5}}} \;
    ((\XZC{0}\cdot\JE{0})_{3}\cdot\PN{1}{0}\,\PM{4}{1}{0})_{2} - \\[1ex]
&&
    \frac{16}{5} \, {\sqrt{\frac{42}{5}}} \;
    ((\XZC{0}\cdot\JE{0})_{2}\cdot{\PN{1}{0}}^3)_{2} +
    16 \, {\sqrt{\frac{6}{5}}} \;
    ((\XZC{0}\cdot\JE{0})_{2}\cdot{\PN{1}{0}}^2\,\PM{2}{1}{0})_{2} - \\[1ex]
&&
    \frac{4}{5} \, {\sqrt{42}} \;
    ((\XZC{0}\cdot\JE{0})_{2}\cdot\PN{1}{0}\,\PM{4}{1}{0})_{2} +
    i \, \frac{48}{5} \, {\sqrt{\frac{14}{5}}} \;
    ((\XZC{0}\cdot\AE)_{2}\cdot{\PN{1}{0}}^2)_{2} - \\[1ex]
&&
    i \, 48 \, {\sqrt{\frac{2}{5}}} \;
    ((\XZC{0}\cdot\AE)_{2}\cdot\PN{1}{0}\,\PM{2}{1}{0})_{2} +
    i \, \frac{12}{5} \, {\sqrt{14}} \;
    ((\XZC{0}\cdot\AE)_{2}\cdot\PM{4}{1}{0})_{2}  \,,  
\\[2ex]
    0  &=& -
    i \, \frac{126}{{\sqrt{5}}} \,
    (\{\ZZC{0},\XZC{0}\}_{4}\cdot{\PN{1}{0}}^2)_{4} +
    i \, 78 \, {\sqrt{2}} \;
    (\{\ZZC{0},\XZC{0}\}_{3}\cdot\PN{1}{0}\,\PM{2}{1}{0})_{4} + \\[1ex]
&&
    i \, 12 \, {\sqrt{385}} \;
    (\{\ZZC{0},\XZC{0}\}_{3}\cdot\PM{4}{1}{0})_{4} -
    i \, 153 \, {\sqrt{\frac{7}{5}}} \;
    (\{\ZZC{0},\XZC{0}\}_{2}\cdot\PN{1}{0}\,\PM{2}{1}{0})_{4} - \\[1ex]
&&
    i \, 35 \, {\sqrt{\frac{11}{2}}} \;
    (\{\ZZC{0},\XZC{0}\}_{2}\cdot\PM{4}{1}{0})_{4} -
    i \, 127 \, (\{\ZZC{0},\XZC{0}\}_{0}\cdot\PM{4}{1}{0})_{4} - \\[1ex]
&&
    i \, \frac{994}{3} \, \sqrt{2} \;
    ((\ZZC{0}\cdot\BE)_{3}\cdot\PN{1}{0}\,\PM{2}{1}{0})_{4} -
    i \, \frac{301}{3} \, {\sqrt{385}} \;
    ((\ZZC{0}\cdot\BE)_{3}\cdot\PM{4}{1}{0})_{4} + \\[1ex]
&&
    i \, \frac{17906}{3} \sqrt{\frac{2}{5}} \;
    ((\ZZC{0}\cdot\BE)_{2}\cdot\PN{1}{0}\,\PM{2}{1}{0})_{4} +
    i \, \frac{679}{6} \, {\sqrt{77}} \;
    ((\ZZC{0}\cdot\BE)_{2}\cdot\PM{4}{1}{0})_{4} + \\[1ex]
&&
    i \, \frac{1925}{2} \, {\sqrt{\frac{5}{3}}} \;
    ((\ZZC{0}\cdot\BE)_{1}\cdot\PM{4}{1}{0})_{4} +
    1024 \, {\sqrt{6}} \;
    ((\XZC{0}\cdot\JE{0})_{3}\cdot{\PN{1}{0}}^2\,\PM{2}{1}{0})_{4} + \\[1ex]
&&
    768 \, {\sqrt{\frac{105}{11}}} \;
    ((\XZC{0}\cdot\JE{0})_{3}\cdot\PN{1}{0}\,\PM{4}{1}{0})_{4} -
    1792 \, {\sqrt{\frac{91}{15}}} \;
    ((\XZC{0}\cdot\JE{0})_{3}\cdot\PM{6}{1}{0})_{4} - \\[1ex]
&&
    \frac{21536}{7} \, {\sqrt{\frac{6}{5}}} \;
    ((\XZC{0}\cdot\JE{0})_{2}\cdot{\PN{1}{0}}^2\,\PM{2}{1}{0})_{4} +
\\[1ex]%
&&    3552 \, {\sqrt{\frac{3}{77}}} \;
    ((\XZC{0}\cdot\JE{0})_{2}\cdot\PN{1}{0}\,\PM{4}{1}{0})_{4} - \\[1ex]
&&
    808 \, {\sqrt{\frac{130}{3}}} \;
    ((\XZC{0}\cdot\JE{0})_{2}\cdot\PM{6}{1}{0})_{4} +
    i \, \frac{14304}{7} \, {\sqrt{\frac{2}{5}}} \;
    ((\XZC{0}\cdot\AE)_{2}\cdot\PN{1}{0}\,\PM{2}{1}{0})_{4} - \\[1ex]
&&
    i \, \frac{60960}{\sqrt{77}} \,
    ((\XZC{0}\cdot\AE)_{2}\cdot\PM{4}{1}{0})_{4} -
    i \, 152 \, {\sqrt{\frac{26}{5}}} \;
    ((\XZC{0}\cdot\AE)_{2}\cdot\frac{\PM{6}{1}{0}}{\PN{1}{0}})_{4}
    \,; 
\end{array}
\end{equation*}
\begin{equation*}
\begin{array}{lcl}
    0  &=&
i \, \frac{1}{{\sqrt{3}}} \:
(\{\ZZC{0},\ZZC{0}\}_{3}\cdot\PM{2}{1}{0})_{2} -
i \, \frac{45}{7} \,
(\{\XZC{0},\XZC{0}\}_{3}\cdot\PN{1}{0}\,\PM{2}{1}{0})_{2} + \\[1ex]
&&
i \, 9 \, {\sqrt{\frac{5}{14}}} \;
(\{\XZC{0},\XZC{0}\}_{3}\cdot\PM{4}{1}{0})_{2} +
i \, 52 \, {\sqrt{\frac{6}{35}}} \;
((\ZZC{0}\cdot\AE)_{2}\cdot\PN{1}{0})_{2} + \\[1ex]
&&
i \, 12 \, {\sqrt{\frac{6}{7}}} \;
((\ZZC{0}\cdot\AE)_{1}\cdot\PM{2}{1}{0})_{2} +
i \, \frac{480}{7} \,
((\XZC{0}\cdot\BE)_{3}\cdot\PN{1}{0}\,\PM{2}{1}{0})_{2} - \\[1ex]
&&
i \, 48 \, {\sqrt{\frac{10}{7}}} \;
((\XZC{0}\cdot\BE)_{3}\cdot\PM{4}{1}{0})_{2} -
i \, 96 \, {\sqrt{\frac{2}{35}}} \;
((\XZC{0}\cdot\BE)_{2}\cdot{\PN{1}{0}}^2)_{2} + \\[1ex]
&&
i \, \frac{96}{7} \, {\sqrt{10}} \;
((\XZC{0}\cdot\BE)_{2}\cdot\PN{1}{0}\,\PM{2}{1}{0})_{2} -
i \, 24 \, {\sqrt{\frac{2}{7}}} \;
((\XZC{0}\cdot\BE)_{2}\cdot\PM{4}{1}{0})_{2} - \\[1ex]
&&
\frac{960}{7} \, {\sqrt{3}} \;
(((\JE{0}\cdot\JE{0})_{2}\cdot\JE{0})_{3}\cdot{\PN{1}{0}}^2\,\PM{2}{1}{0})_{2} + \\
&&
96 \, {\sqrt{\frac{30}{7}}} \;
(((\JE{0}\cdot\JE{0})_{2}\cdot\JE{0})_{3}\cdot\PN{1}{0}\,\PM{4}{1}{0})_{2} \,.
\end{array}
\end{equation*}
%
%
%
In the sequel we shall refer to the right hand sides of the
preceeding constraint equations as DYN1, DYN2, DYN3, and DYN0
respectively.

10 [!] $P$-linear independent combinations of the components of
the Poisson brackets $\{ \ZZC{0} , \XZC{0} \}_L$, $L=2,3$, remain
unconstrained. A convenient choice of these combinations consists
in the 5 components of the irreducible tensor $\{ \ZZC{0} , \XZC{0}
\}_2$ and 5 $P$--linearly independent combinations of the components
of the irreducible tensor $\{ \ZZC{0} , \XZC{0} \}_3$ which are not
constrained by DYN2 = 0.

11 [!] $P$-linear independent combinations of the components of
the Poisson brackets $\{ \XZC{0} , \XZC{0} \}_L$ and $\{ \ZZC{0}
, \ZZC{0} \}_L$ remain unconstrained. A convenient choice of these
combinations consists in the 3 + 7 components of the irreducible
tensors $\{ \XZC{0} , \XZC{0} \}_L$, $L=1,3$ and the ``projection''
$( \{ \ZZC{0} , \ZZC{0} \}_3 \cdot \PM{3}{1}{0} )_0$.

The derivation of the above dynamical constraints does not require
any specifications of the ``mixed'' Poisson brackets beyond the
ones already stated in the context of the kinematical constraints.
If for $\XZC{0}$ and $\ZZC{0}$ the right hand sides of their
composition laws are inserted and the (isomorphic) generalized
Poisson commutation relations for the merging branches are used,
the dynamical constraints are identically satisfied. Thus, the
concrete specifications for the ``mixed'' Poisson brackets, given
so far, pass the test of consistency with the generalized Poisson
commutation relations for the three branches.

\bs
Comparing the numbers for the independent combinations of the
components $\{ \ZZC{0} , \XZC{0} \}_{L,M}$; $\{ \XZC{0} , \XZC{0}
\}_{L,M}$ and $\{ \ZZC{0} , \ZZC{0} \}_{L,M}$ which remain
unconstrained, on the one hand, with the previously determined
numbers for the independent
new relations (of the restricted type) among the basis elements
of the subspaces $\GV^{[k]l}$, $k=0,1,2$, $0 \le l \le 2$, on the
other hand, we conclude that the latter relations can all be
obtained from the composition laws $\XZC{0} = \dots$ and $\ZZC{0}
= \dots$ by Poisson bracket induction. Moreover, produced in this
way they are automatically organized in the irreducible tensor
multiplets:
\begin{align*}
&  \{ \ZZC{0} , \XZC{0} \}_2 = \dots ,\\[2mm]
&  \{ \ZZC{0} , \XZC{0} \}_3 \Big/_{\text{DYN2 = 0}}  =
    \dots\dots\dots \Big/_{\text{DYN2RHS = 0}}   ;\\[2mm]
&  \{ \XZC{0} , \XZC{0} \}_L = \dots ,\qquad\qquad L=1,3\\[2mm]
&   ( \{ \ZZC{0} , \ZZC{0} \}_3 \cdot \PM{3}{1}{0} )_0 = \dots .
\end{align*}
Here the dots and the symbol DYN2RHS denote the result of
inserting the right hand sides of the composition laws
$\XZC{0} = \dots$ and $\ZZC{0} = \dots$ into the Poisson
brackets and into the irreducible tensor DYN2, respectively,
on the pertinent left hand sides.
\es

Just for the record: these manifestly covariant ``new'' composition laws
are new only as far as  they cannot be produced by {\it multiplication}
from relations among the basis elements of the subspaces $\GV^{[k]l}$,
$k=0,1,2$, $0 \le l \le 1$.

\vspace{4mm}
Before we turn to the quantization of the composition laws
$\XZC{0} = \dots$ and $\ZZC{0} = \dots$,  we note in passing
that the explicit expressions for the generators $\TZ{0}$ and
$\SZ{0}$ in terms of matrix elements of the relevant monodromy
matrix suggest the following specifications for the Poisson brackets
\[ (i)   \quad  \{ \TZ{0} , \PE{1}{0} \}_2 = 0 , \qquad
   (ii)  \quad  \{ \TZ{0} , \PE{1}{0} \}_3 = 0 , \qquad \text{and}\qquad
   (iii) \quad  \{ \SZ{0} , \PE{1}{0} \}_3 = 0 .\]
This observation is based on the fact that the first entries
$T^{[0]}_{2,m_1}$ ($S^{[0]}_{2,m_1}$) in the Poisson brackets
of the non-vanishing contributions
\[  \langle 2,m_1 ; 1,m_2 | L,M \rangle \:
    \{  T^{[0]}_{2,m_1} , P^{[1]}_{1,m_2 | 0} \}  \qquad
    ( \langle 2,m_1 ; 1,m_2 | L,M \rangle \:
    \{  S^{[0]}_{2,m_1} , P^{[1]}_{1,m_2 | 0} \} )  \]
to the left hand sides of $(i)$ \& $(ii)$ and of $(iii)$, respectively,
do not contain dynamical variables which are canonically
conjugate to the dynamical variables contained in the
corresponding second entries $P^{[1]}_{1,m_2 | 0}$.

For a similar reason we expect that an evaluation of the
Poisson brackets of $\SZ{0}$ and $\SE{0}$ with $\pominus$
would result in $P$--linear combinations of the components
$J^{[0]}_{1,m}$.

The importance of these observations for the quantization
of the composition laws (\ref{comp1}) and (\ref{comp2})
will become clear in the next section.

\newcommand{\ghh}{\hat{\mathfrak{h}}}
\newcommand{\GVH}{\hat{\mathfrak{V}} }

\newcommand{\HB}{\hat{B}}
\newcommand{\HJ}{{\hat{J}}}
\newcommand{\HS}{\hat{S}}
\newcommand{\HT}{\hat{T}}

\newcommand{\HW}{\hat{W}}
\newcommand{\HAE}{\hat{\mathscr{A}}}
\newcommand{\HBE}{\hat{\mathscr{B}}}

\newcommand{\HJQ}{\HJ \!\!\!   \backslash}
\newcommand{\HBQ}{\HB \!\!\!   \backslash}
\newcommand{\HSQ}{\HS \!\!\!   \backslash}
\newcommand{\HTQ}{\HT \!\!\!\! \backslash}

\newcommand{\HAEQ}{\hat{\mathscr{A}} \!\!\!\! \backslash}
\newcommand{\HBEQ}{\hat{\mathscr{B}} \!\!\!\! \backslash}

\newcommand{\HSO}{\hat{\mathscr{O}}}
\newcommand{\HSOQ}{\hat{\mathscr{O}} \!\!\!\! \backslash}

\renewcommand{\PE}[2]{\hat{P}_{1|#2}^{[#1]}}
\renewcommand{\PN}[2]{(\hat{P}_{1|#2}^{[#1] 2})_{0}}
\renewcommand{\PM}[3]{(\hat{P}_{1|#3}^{[#2] #1})_{#1}}

\renewcommand{\PO}[1]{\hat{P}_{0|0}^{[#1]}}

\renewcommand{\WE}[2]{\HW_{1|#2}^{[#1]}}
\renewcommand{\AE}{{\HAE}_{1|0}}
\renewcommand{\BE}{{\HBE}_{1|0}}

\renewcommand{\se}{\hat{s}_1}
\renewcommand{\te}{\hat{t}_1}


\section*{Quantum Theory}

In this section we explore the chances that the previously presented
classical composition laws can be quantized and we narrow down the
shape they would take in the quantum theory.

To shed light on these issues, we stick to the strategy of Ref.\
\cite{KP 99} :
\\
We make use of the 1:1 correspondence established there between
the classical observables $\JE{k}$, $\BNE{k}$, $\SE{k}$, $\SZ{k}$,
$\TZ{k},\dots$ on the one hand, and the quantum observables
\renewcommand{\JE}[1]{{\HJQ_1^{[#1]}}}
\renewcommand{\BNE}[1]{\HBQ_{0}^{(1)[#1]}}
\renewcommand{\TZ}[1]{\HTQ_2^{\,[#1]}}
\renewcommand{\SZ}[1]{\HSQ_2^{[#1]}}
\renewcommand{\SE}[1]{\HSQ_1^{[#1]}}
\[
\hat{J}^{[0]}_1 = \frac{\hbar}{2\pi\alpha'} \, \JE{k},\qquad
\hat{B}^{(1)[0]}_0 = (\frac{\hbar}{2\pi\alpha'})^2\,\BNE{k},\qquad
\hat{S}^{[0]}_2 = (\frac{\hbar}{2\pi\alpha'})^2\,\SE{k},
\]
\[
\hat{S}^{[0]}_1 = (\frac{\hbar}{2\pi\alpha'})^2\,\SZ{k},\qquad
\hat{T}^{[0]}_2 = (\frac{\hbar}{2\pi\alpha'})^2\,\TZ{k},
\qquad\dots\qquad\qquad  k=0,1,2
\]
on the other hand.

Here for each separate value of $k$, the ``slashed'' quantities act
as hermitian, scale invariant generators of quantum $*$--algebras
$\ghh^{[k]-}$ which correspond to the classical $*$--algebras $\gh^{[k]-}$,
$k=0,1,2$. The validity of the generalized commutation relations for
each branch separately is left untouched. The elements of the algebras
$\ghh^{[1]-}$ and $\ghh^{[2]-}$ commute with one another for reasons
of causality as before. The quantum counterparts $\hat{P}_\mu^{[k]}$
of the components of the energy momentum vectors $P_\mu^{[k]}$ act
as central elements for each algebra $\gh^{[k]-}$, $k=0,1,2$, separately.
Moreover, the quantum counterparts $\hat{P}_\mu^{[0]}$ act as global
central elements for all three algebras $\gh^{[k]-}$, $k=0,1,2$, and
beyond for all quantum observables of the string vertex built from the
right-movers. Hence, without loss of generality, the components
$\hat{P}_\mu^{[0]}$, in particular the (total mass)$^2$--operator
$(\hat{P}^{[0]})^2 = \hat{P}_\mu^{[0]} \hat{P}^{[0]\mu}$, may be
treated as $c$-numbers and may be fixed. Since we are primarily
interested in vertices describing real processes with a positive
mass $\mn$ for the merged branch, we fix the components of
$\hat{P}_\mu^{[0]}$, $\mu = 0,1,2,3$, as numbers in the forward
light-cone in momentum space. Due to Lorentz invariance, we may
even set
\[  \hat{P}_\mu^{[0]} = \delta_\mu^0 \, \mn ,\qquad\qquad\qquad
    \mn > 0 .\]
This defines the (class of) rest frame(s) for the merged branch.
Energy--momentum conservation requires the following relations
for the branch [0] rest frame components
$\PO{k} := (\hat{P}_\mu^{[k]} \hat{P}^{[0]\mu})/\mn$
and $\hat{P}_{n|0}^{[k]}$, $k=1,2$: $\PO{1} + \PO{2} = \mn$ and
$\hat{P}_{n|0}^{[1]} + \hat{P}_{n|0}^{[2]} = 0$, $n=1,2,3$, or
rather $\hat{P}_{1|0}^{[1]} = - \hat{P}_{1|0}^{[2]}$ referring
to a branch [0] rest frame spin basis.

The components of the three operator-valued vectors $\hat{P}_\mu
^{[k]}$, $\mu = 0,1,2,3$, can be simultaneously diagonalized.
Hence, for the real vertex processes we have in mind, it is
legitimate to limit our angle of sight of the entire algebra
to the perspective from its momentum type representations,
for which $\hat{P}_{1,m|0}^{[1]}$, $m=-1,0,+1$ and $(\PO{1}
- \PO{2} )$ act as multiplication operators. Thus we take the
liberty of dropping the hats from the $\hat{P}_{..}^{[\cdot]}$
operators, at the same same time paying utmost attention to
the order in which the latter ones appear when combined with
non-commuting partners.\\
We postulate the validity of the following ``mixed'' commutation
relations
\renewcommand{\PE}[2]{P_{1|#2}^{[#1]}\rule{0ex}{2ex}}
\renewcommand{\PN}[2]{(P_{1|#2}^{[#1] 2})_{0}}
\renewcommand{\PM}[3]{(P_{1|#3}^{[#2] #1})_{#1}}
\renewcommand{\PO}[1]{P_{0|0}^{[#1]}}
\renewcommand{\BNE}[1]{\HB_{0}^{(1)[#1]}}
\renewcommand{\TZ}[1]{\HT_2^{\,[#1]}}
\renewcommand{\SZ}[1]{\HS_2^{[#1]}}
\renewcommand{\SE}[1]{\HS_1^{[#1]}}
\[  [ \JE{0} , \HSO_j ]_L = - \delta_{L,j} \: \sqrt{j(j+1)} \;
    \HSO_j  ,\qquad\qquad\qquad j=0,1,2,\dots  \]
for all irreducible tensor operators $\HSO_j$ contained in $\ghh^{[k]}$,
$k=0,1,2$, in particular:
\[  [ \JE{0} , \PE{1}{0} ]_L = - \delta_{L,1} \: \sqrt{2} \; \PE{1}{0},
    \qquad\qquad
    [ \JE{0} , \pominus ]_1 = 0 . \]
Further:
\[  [ \TZ{0} , \PE{1}{0} ]_L = 0,
    \qquad L=2,3 \qquad\quad\text{and}\qquad\quad
    [ \SZ{0} , \PE{1}{0} ]_3 = 0  .\]
Faithful to the correspondence principle as our guide, we postulate
that
\renewcommand{\JE}[1]{{{\HJ}_1^{[#1]}\rule{0ex}{2ex}}}
\begin{itemize}
\item[{\it i})]
there does not exist any quantum observable which -- as far as its
mass dimensions are concerned -- scales like a negative power of
$(\frac{\hbar}{2\pi\alpha'})$;
\item[{\it ii})]
$\frac1{(\hbar / 2\pi\alpha')}$ times the commutator of two
quantum observables produces a quantum observable or zero;
\item[{\it iii})]
there exists a one to one correspondence between the relations
among the classical observables contained in the classical
$*$--algebras $\gh^{[k]-}$, $k=0,1,2$, and the relations
among the quantum observables contained in the quantum
$*$--algebras $\ghh^{[k]-}$, $k=0,1,2$.
\end{itemize}
Up to quantum corrections, the quantum relations can be obtained
from the classical relations according to the following rules:
\begin{itemize}
\item[$\alpha$)]
replace each (simple) Poisson bracket $\{ \cdot , \cdot \}_j$ of
the classical irreducible tensor variables by $(\frac{i \hbar}
{2\pi\alpha'})^{-1}$ times the commutator $[ \cdot , \cdot ]_j$
of their respective (original) counterparts in $\ghh^{[k]-}$,
$k=0,1,2$, without changing the original succession of brackets
inside an iterated bracket;
\item[$\beta$)]
replace each (simple) product $( \cdot , \cdot )_j$ of the
classical irreducible tensor variables by $\frac12$ times the
anticommutator $[ \cdot , \cdot ]^+_j$ of their respective
(original) counterparts in $\ghh^{[k]-}$, $k=0,1,2$, without
changing the original succession of brackets inside an iterated
bracket.
\end{itemize}
Now that the Poisson bracket action has been replaced by the
commutator action $[ \cdot , \cdot ]_j$, the symbol $\{ \cdot ,
\cdot \}_j$ is
available anew. From now on, we shall employ the curly brackets
$\{ \cdot , \cdot \}_j$ (rather than the straight brackets $[ \cdot ,
\cdot ]^+_j$) as the symbol for the anticommutator.

The quantum corrections are distinguished from the so-called
classical parts of the quantum relations (just constructed)
in that the coefficients of the irreducible tensor operators
appearing in these corrections display explicit positive integer
powers in $(\frac{\hbar}{2\pi\alpha'})$.

The above replacement of products by anticommutators has
significant advantages over alternative replacements because
of the relative simplicity of the recoupling formulae for
commutators, anticommutators and a mix of both involving
three irreducible tensor operators, \cf\ Ref.\ \cite{KP 99}.

Moreover, manipulations of operator-valued dynamical variables
in terms of commutators and anticommutators differ from those
of the corresponding classical dynamical variables in terms of
Poisson brackets and anticommutators by {\it double} commutators
only (\cf\ Ref.\ \cite{KP 99}).

The quantization of the composition law $(\mathscr{A}_{1|0} \cdot
\PE{1}{0} )_0 = 0$ does not pose any problems. In quantum theory
it assumes the shape
\[  \{  \AE , \PE{1}{0}  \}_0  = 0 \]
with
\beas  \AE &:=& - i \, \JE{0} \, \potimes +
                 \WE{1}{0} \, \poz + \WE{2}{0} \, \poe  \, ,\\
       \BE &:=& - i \, \JE{0} \, \pominus + \WE{1}{0} - \WE{2}{0}  \,,
\eeas
where
\[  \WE{k}{0} := i \, \big[ \JE{k} \, \gm^{[k]} +
                 {\textstyle \frac12} \: \{ \JE{k} , \PE{\neq k}{0} \}_0 \,
                 \delta^{[k]} \, \PE{k}{0} \big] .\]
In order to explore the chances for a consistent quantization of
the composition laws (\ref{comp1}) and (\ref{comp2}) and to narrow
down their likely appearances as operator relations, we resort to
the scrutiny of all conceivable preconditions they have to satisfy
in quantum theory.

Apart from the requirements related to covariance, hermiticity,
exchange (anti)symmetry and to the correspondence principle,
these preconditions include the observance of
\begin{itemize}
\item[{\it a})]
(an adequate quantal version of) the kinematical constraints,
\item[{\it b})]
(an adequate quantal version of) the dynamical constraints.
\end{itemize}
The requirements related to covariance (both with respect to
the Lorentz and to the $\GV^{[0]}$ algebras, respectively),
hermiticity, exchange (anti)symmetry and to the correspondence
principle for both sides of the composition laws are easily
met (see above):
\\
starting with the right hand sides, the r.h.s.\
of (\ref{comp1}) goes over into
\[
    \Big[ \hat\Xi^{[1]}_{2|0} - \hat\Xi^{[2]}_{2|0} \Big] \,
{\textstyle \frac1{\mn}} -
           4 \, (\AE^2)_2 \:
           \Big\langle  {\textstyle \frac{\pominus}{\potimesq}} \Big\rangle +
           4 \, \{ \AE , \BE \}_2
           \Big\langle {\textstyle \frac{1}{\potimes}} \Big\rangle + \hspace{4.8em}\]
\[  \hspace{14.5em}\hq^2   \, \PM{2}{1}{0} \,
    \Big\langle  {\textstyle \frac{\pominus}{\mnq}} \:
    \xi  \Big( \Big( {\textstyle \frac{\pominus}{\mn}} \Big)^2 ,
    {\textstyle \frac{\PN{1}{0}}{\mnq}} \Big) \Big\rangle  ,\]
the r.h.s.\ of (\ref{comp2}) goes over into
\[  \Big[ \hat\Theta^{[1]}_{2|0} + \hat\Theta^{[2]}_{2|0} \Big] \,
    {\textstyle \frac1{\mn}} +
    12 \, (\AE^2)_2
    \Big\langle {\textstyle \frac{1}{\potimes}}  \Big\rangle +
    \hq^2 \, \PM{2}{1}{0} \,
    \Big\langle  \vartheta  \Big( \Big( 
    {\textstyle \frac{\pominus}{\mn}} \Big)^2 ,
    {\textstyle \frac{\PN{1}{0}}{\mnq}} \Big) \Big\rangle  .\]
Here and in the sequel the symbol $\hq$ stands for the fraction
$\frac{\hbar}{2 \pi \alpha'}$.
The above irreducible tensor operators $\hat\Xi^{[k]}_{2|0}$ and
$\hat\Theta^{[k]}_{2|0}$, $k=1,2$, are obtained from the irreducible
tensor variables $\Xi^{[k]}_{2|0}$ and $\Theta^{[k]}_{2|0}$,
respectively, by replacing
\beas
    \XZC{k}  &\text{by}&
    \XZQ{k}  =  - \imag \,{\textstyle \sqrt{\frac32}}\;
                \{ \SZ{k} , \PE{\neq k}{k} \}_{2} + {\textstyle \frac12} \,
                \{ \SE{k} , \PE{\neq k}{k} \}_{2} , \\
    \ZZC{k}  &\text{by}&
    \ZZQ{k}  = - {\sqrt{3}} \; \{ \TZ{k} , \PN{\neq k}{k} \}_{2} +
                 {\sqrt{21}}\; \{ \TZ{k} , \PM{2}{\neq k}{k})_{2} -
                 {\textstyle \frac12} \, \{ \BNE{k} , \PM{2}{\neq k}{k} \}_{2}  ,
\eeas
$k=1,2$, $( \mathscr{A}_{1|0}^2 )_2$ by $( \AE^2 )_2$ and $(
\mathscr{A}_{1|0} \cdot \mathscr{B}_{1|0} )_2$ by $\frac12 \,
\{ \AE , \BE \}_2$.\\
$\xi( ( \frac{\pominus}{\mn} )^2 , \frac{\PN{1}{0}}{\mnq} )$ and
$\vartheta( ( \frac{\pominus}{\mn} )^2 , \frac{\PN{1}{0}}{\mnq} )$
are scalar, real, as yet undetermined numerical functions of their
dimensionless arguments, functions which are symmetric under the
exchange of the branches [1] and [2].

Turning to the left hand sides of the composition laws (\ref{comp1})
and (\ref{comp2}), or rather to the equations defining them, the
symbol $\XZC{0}$ on the l.h.s.\ of (\ref{comp1}) is defined by
\[  \XZQ{0} :=  - \imag \,{\textstyle {\sqrt{\frac32}}}\;
                \{ \SZ{0} , \PE{1}{0} \}_{2} + {\textstyle \frac12} \,
                \{ \SE{0} , \PE{1}{0} \}_{2}  ,\]
while the symbol $\ZZC{0}$ on the l.h.s.\ of (\ref{comp2}) is
replaced by
\[  \ZZQ{0}  :=
       {\textstyle \frac94} \, \{\{\TZ{0},\PE{1}{0}\}_{2},\PE{1}{0}\}_{2} +
       {\textstyle \frac14} \, {\sqrt{15}} \;
       \{\{\TZ{0},\PE{1}{0}\}_{1},\PE{1}{0}\}_{2} -
       {\textstyle \frac14} \, \{\{\BNE{0},\PE{1}{0}\}_{1},\PE{1}{0}\}_{2}  \,.\]
This way of defining $\ZZQ{0}$ ensures that the kinematical
constraint for the composition law (\ref{comp2}) is
{\it identically} satisfied for both sides of (\ref{comp2})
in the form:
\[  \{ \{ \ZZQ{0} , \PE{1}{0} \}_{1} , \PE{1}{0}\}_{1}  \; = \; 0 , \]
%
\begin{equation*}
\begin{array}{c}
  \{ \{ \frac1{\mn} \,
    \Big[ \hat\Theta^{[1]}_{2|0} + \hat\Theta^{[2]}_{2|0} \Big] +
    \frac{12}{\potimes} \, (\AE^2)_2  +
    \hq^2 \, \PM{2}{1}{0} \,
    \Big\langle  \vartheta  \Big( \Big( \frac{\pominus}{\mn} \Big)^2 ,
    \frac{\PN{1}{0}}{\mnq} \Big) \Big\rangle  , 
    \hspace{2cm}\\
%
    \hspace{11.7cm}
      \PE{1}{0} \}_{1} , \PE{1}{0}\}_{1}  \; =  \; 0 . 
\end{array}
\end{equation*}
%
(This is also true, if the intermediate spin-channel index 1 is
replaced by the intermediate spin-channel index 2.)

As before, we define the commutators $[ \XZQ{0} , \PE{1}{0} ]_L$,
$[ \ZZQ{0} , \PE{1}{0} ]_L$, $L=1,2,3$, $[ \XZQ{0} , \pominus ]_2$
and $[ \ZZQ{0} , \pominus ]_2$ in accordance with the right hand
sides of their composition laws. In this way, we are led to specifications
of the commutators just mentioned which literally agree with the
application of substitution rules $\alpha$) and $\beta$) to the
specifications of the corresponding Poisson brackets.

Next, we replace the subset of generators of $\ghh^{[0]-}$:
$\BNE{0}$, $\SE{0}$, $\SZ{0}$, $\TZ{0}$ by the following
equivalent subset of irreducible tensor operators:
\[  \se := \{ \SZ{0} , \PE{1}{0} \}_1  \quad,\quad \XZQ{0} ;\qquad\qquad\quad
    \te := \{ \TZ{0} , \PE{1}{0} \}_1  \quad,\quad \ZZQ{0}  .\]
Further, using the formal properties of the commutators, we convert
our various specifications of the ``mixed'' commutators into the
equivalent form
\begin{align*}
\CZ := [ \se , \PE{1}{0} ]_2 &=
{\textstyle\frac{1}{14}} \, \{\CE,{\textstyle\frac{\PM{2}{1}{0}}{\PN{1}{0}}}\}_{2} +
{\textstyle\frac{1}{20}} \, \{\CN,{\textstyle\frac{\PM{2}{1}{0}}{\PN{1}{0}}}\}_{2} - \\ &\quad\;\,
{\textstyle\frac{24}{35}} \, {\textstyle{\sqrt{\frac{3}{5}}}} \; \hq  
\,\{\BE,\PE{1}{0}\}_{2} +
{\textstyle\frac{8}{5\,{\sqrt{7}}}} \; \hq 
\, \{\BE,{\textstyle\frac{\PM{3}{1}{0}}{\PN{1}{0}}}\}_{2}
 \,, \\
\CE := [ \se , \PE{1}{0} ]_1   &  \,, \\
\CN := [ \se , \PE{1}{0} ]_0   &  \,,
\end{align*}
\[  [ \XZQ{0},\PE{1}{0} ]_L = \begin{cases}
    i \, 2 \,{{\textstyle\sqrt{\frac{10}{3}}}}\; \hq \,\{\BE , \PE{1}{0}\}_{1}
    \qquad\qquad &
    \text{for $L=1$} \\
    i \, 2 \,{\sqrt{6}}\; \hq \,\{\BE , \PE{1}{0}\}_{2} &
    \text{for $L=2$} \\
    0 & \text{for $L=3$}  \,, \end{cases}   \]
\begin{align*}
\KZ := [ \te , \PE{1}{0} ]_2 &=
- i \, {{\textstyle\sqrt{\frac{2}{15}}}} \; \hq \,
\{\AE,{\textstyle\frac{\PM{2}{1}{0}}{\PN{1}{0}}}\}_{2} +
{\textstyle\frac{1}{20}} \: \{\KN,{\textstyle\frac{\PM{2}{1}{0}}{\PN{1}{0}}}\}_{2}  \,,\\
\KE := [ \te , \PE{1}{0} ]_1 &=
i \, 2 \, {\textstyle{\sqrt{\frac{10}{3}}}} \;  \hq \,\AE  \,, \\
\KN := [ \te , \PE{1}{0} ]_0    &  \,,
\end{align*}
\[  [ \ZZQ{0},\PE{1}{0} ]_L = \begin{cases}
    i \, 2 \, {\sqrt{30}} \;  \hq \, \{\AE , \PE{1}{0}\}_{1} \qquad\qquad  &
    \text{for $L=1$} \\
    i \, 6 \, {\sqrt{6}} \; \hq \, \{\AE , \PE{1}{0}\}_{2}  &
    \text{for $L=2$} \\
    0 & \text{for $L=3$}  \,. \end{cases}    \]
With the help of the above definitions of $\CE$, $\CN$ and $\KN$,
we can express the original  generators in terms
of $\se$, $\XZQ{0}$; $\te$, $\ZZQ{0}$:
%
\begin{equation*}
    \begin{array}{lcl}
\SE{0} &=&  i \, \frac{1}{3} \, {\sqrt{\frac{5}{2}}} \;
            \{\se,\frac{\PE{1}{0}}{\PN{1}{0}}\}_{1} +
            \frac{3}{2\,{\sqrt{5}}} \:
            \{\XZQ{0},\frac{\PE{1}{0}}{\PN{1}{0}}\}_{1} -
            \frac{1}{12} \, {\sqrt{\frac{7}{3}}} \;
            \{\XZQ{0},\frac{\PM{3}{1}{0}}{{\PN{1}{0}}^2}\}_{1} + \\
       & &  i \, \frac{43}{630 \, {\sqrt{2}}} \;
            \{[\CE,\PE{1}{0}]_{2},\frac{\PE{1}{0}}{{\PN{1}{0}}^2}\}_{1} +
            i \, \frac{11}{9} \, {\sqrt{\frac{2}{105}}} \;
            \{[\CE,\PE{1}{0}]_{2},\frac{\PM{3}{1}{0}}{{\PN{1}{0}}^3}\}_{1} -\\
       & &  i \, \frac{5}{63} \, {\sqrt{\frac{5}{6}}} \;
            \{[\CE,\PE{1}{0}]_{1},\frac{\PE{1}{0}}{{\PN{1}{0}}^2}\}_{1} +
            i \, \frac{7}{180 \, {\sqrt{10}}} \;
            \{[\CN,\PE{1}{0}]_{1},\frac{\PE{1}{0}}{{\PN{1}{0}}^2}\}_{1} + \\
       & &  \sqrt{3} \;  \hq^2 \,\PE{1}{0} \: \frac{\pominus}{\PN{1}{0}}
            \,, 
\\
%
\SZ{0} &=&  \frac{3}{4\,{\sqrt{5}}} \:
            \{\se,\frac{\PE{1}{0}}{\PN{1}{0}}\}_{2} -
            \frac{1}{24} \, {\sqrt{\frac{7}{3}}} \;
            \{\se,\frac{\PM{3}{1}{0}}{{\PN{1}{0}}^2}\}_{2} -
            i \, \frac{1}{5 \, {\sqrt{2}}} \:
            \{\XZQ{0},\frac{\PE{1}{0}}{\PN{1}{0}}\}_{2} + \\
       & &  i \, \frac{1}{12} \, {\sqrt{\frac{7}{5}}} \;
            \{\XZQ{0},\frac{\PM{3}{1}{0}}{{\PN{1}{0}}^2}\}_{2} -
            \frac{3}{560 \, \sqrt{5}} \:
            \{[\CE,\PE{1}{0}]_{2},\frac{\PE{1}{0}}{{\PN{1}{0}}^2}\}_{2} + \\
       & &  \frac{257}{720 \, \sqrt{14}} \;
            \{[\CE,\PE{1}{0}]_{2},\frac{\PM{3}{1}{0}}{{\PN{1}{0}}^3}\}_{2} -
            \frac{11}{280} \, {\sqrt{\frac{3}{5}}} \;
            \{[\CE,\PE{1}{0}]_{1},\frac{\PE{1}{0}}{{\PN{1}{0}}^2}\}_{2} -\\
       & &  \frac{73}{1440 \, \sqrt{7}} \;
            \{[\CE,\PE{1}{0}]_{1},\frac{\PM{3}{1}{0}}{{\PN{1}{0}}^3}\}_{2} +
            \frac{57}{800 \, \sqrt{5}} \;
            \{[\CN,\PE{1}{0}]_{1},\frac{\PE{1}{0}}{{\PN{1}{0}}^2}\}_{2} -\\
       & &  \frac{409}{4800} \,{\sqrt{\frac{7}{3}}} \;
            \{[\CN,\PE{1}{0}]_{1},\frac{\PM{3}{1}{0}}{{\PN{1}{0}}^3}\}_{2} \,,
%
\\
\BNE{0} &=& \frac{{\sqrt{15}}}{4} \:
            \{\te,\frac{\PE{1}{0}}{\PN{1}{0}}\}_{0} -
            \frac{{\sqrt{5}}}{4} \:
            \{\ZZQ{0},\frac{\PM{2}{1}{0}}{{\PN{1}{0}}^2}\}_{0} + \\
       & &  \frac{11}{80} \, {\sqrt{\frac{3}{5}}} \;
            \{[\KN,\PE{1}{0}]_{1},\frac{\PE{1}{0}}{{\PN{1}{0}}^2}\}_{0} +
            \frac{16}{5 \, \sqrt{3}} \:  \hq^2 \, \PN{1}{0} \:
            \frac{\potimes}{{\PN{1}{0}}^2}  \,, 
\\
%
\TZ{0} \; \; \: &=&  \frac{3}{4\,{\sqrt{5}}} \:
            \{\te,\frac{\PE{1}{0}}{\PN{1}{0}}\}_{2} -
            \frac{1}{24} \, {\sqrt{\frac{7}{3}}} \;
            \{\te,\frac{\PM{3}{1}{0}}{{\PN{1}{0}}^2}\}_{2} -
            \frac{1}{24\,{\sqrt{3}}} \:
            \{\ZZQ{0},\frac{1}{\PN{1}{0}}\}_{2} + \\
       & &  \frac{1}{48} \, {\sqrt{\frac{7}{3}}} \;
            \{\ZZQ{0},\frac{\PM{2}{1}{0}}{{\PN{1}{0}}^2}\}_{2} +
            \frac{57}{800 \, \sqrt{5}} \:
            \{[\KN,\PE{1}{0}]_{1},\frac{\PE{1}{0}}{{\PN{1}{0}}^2}\}_{2} - \\
       & &  \frac{409}{4800} \, {\sqrt{\frac{7}{3}}} \;
            \{[\KN,\PE{1}{0}]_{1},\frac{\PM{3}{1}{0}}{{\PN{1}{0}}^3}\}_{2} +
            \frac{19}{15} \, \hq^2 \, \PM{2}{1}{0} \:
            \frac{\potimes}{{\PN{1}{0}}^2} \,.
\end{array}
\end{equation*}
Formally, the errors committed in the course of the corresponding
manipulations involve operators scaling like $\hq^{-2}$. Such
operators do not represent observables. Hence,
the errors are just spurious.

We are now ready to tackle the last and decisive issue, the
clarification of the question whether or not the
quantal composition laws $\XZQ{0} = \dots$ and $\ZZQ{0} = \dots$
are in agreement with the generalized commutation relations
\[ \begin{array}{lcl}
[\TZ{k},\TZ{k}]_{3} &=&
           - i\,[\SZ{k},\SE{k}]_{3} 
           - 8\,\hq\,\{ \JE{k} , ({{\JE{k} }}^{\,2})_{2} \}_3 , \\[2mm]
[\SZ{k},\SZ{k}]_{3} &=&
             i\,2\,[\SZ{k},\SE{k}]_{3} 
           + 4\,\hq\,\{\JE{k},\TZ{k}\}_{3} 
           + 24\,\hq\,\{ \JE{k} , ({\JE{k}}^{\,2})_{2} \}_3, \\[2mm]
[\SZ{k},\SZ{k}]_{1} &=&
           - i\,{\sqrt{{\frac{2}{3}}}}\,[\SZ{k},\SE{k}]_{1} 
           - {\frac{1}{6}}\,{\sqrt{5}}\,[\SE{k},\SE{k}]_{1} +\\[2mm]
&&
             8\,{\sqrt{{\frac{2}{3}}}}\,\hq\,\{\JE{k},\TZ{k}\}_{1} 
           + 16\,{\sqrt{{\frac{2}{15}}}}\,\hq\,
             \{\JE{k},({\JE{k}}^{\,2})_{0}\}_{1} 
           - \hq^3 \, f\, {\sqrt{10}}\,\JE{k} , \\[2mm]
[{\BNE{k}},\TZ{k}]_{2} &=&
             i\,{\sqrt{6}}\,[\SZ{k},\SE{k}]_{2}, \\[2mm]
[\TZ{k},\SZ{k}]_{4} &=& 0, \\[2mm]
[\TZ{k},\SZ{k}]_{3} &=&
             i\,[\TZ{k},\SE{k}]_{3} 
           - 4\,\hq\,\{\JE{k},\SZ{k}\}_{3}, \\[2mm]
[\TZ{k},\SZ{k}]_{2} &=&
           - i \, {\frac{1}{3}}\,{\sqrt{{\frac{7}{2}}}}\, [\TZ{k},\SE{k}]_{2}
           + {\frac{2}{3}}\,{\sqrt{14}}\;\hq\,\{\JE{k},\SZ{k}\}_{2}, \\[2mm]
[ {\BNE{k}},\SZ{k} ]_{2} &=&
    - i\,2\,{\sqrt{{\frac{2}{3}}}}\,[\TZ{k},\SE{k}]_{2}
    + 2\,{\sqrt{{\frac{2}{3}}}}\,\hq\,\{\JE{k},\SZ{k}\}_{2}
    + i\,6\,\hq\,\{\JE{k},\SE{k}\}_{2}, \\[2mm]
[ \BNE{k},\SE{k} ]_{1} &=&
    - i\,6\,{\sqrt{{\frac{2}{5}}}}\,[\TZ{k},\SZ{k}]_{1}
    + 2\,{\sqrt{{\frac{3}{5}}}}\,[\TZ{k},\SE{k}]_{1} - \\[2mm]
&&
      i\,12\,{\sqrt{{\frac{3}{5}}}}\,\hq\,\{\JE{k},\SZ{k}\}_{1}
    - 6\,{\sqrt{2}}\,\hq\,\{\JE{k},\SE{k}\}_{1},
\end{array} \]
$f$ being a free rational common parameter, $k=0,1,2$. In other words,
do the two alternative procedures described below give concordant
results with (or without) the help of suitable choices of the functions
$\xi$ and $\vartheta$?\\
 The two alternative procedures consist of a
separate and a common part: Replace in the expressions DYN0, $\dots$,
DYN3 in turns the tensor variables $\XZC{0}$ and $\ZZC{0}$
\begin{itemize}
\item[{\it a})]
by their respective quantal counterparts $\XZQ{0}$ and $\ZZQ{0}$
\item[{\it b})]
by the right hand side of the respective quantal composition laws
$\XZQ{0} = \dots$ and $\ZZQ{0} = \dots$.
\end{itemize}
Subsequently apply the substitution rules $\alpha$) and $\beta$)
and evaluate the commutators with the help of the generalized
commutation relations. Finally, use the recoupling formulae of
Ref.\ \cite{KP 99} for a suitable presentation of the respective results.

The procedures {\it a}) and {\it b}) involve among other things
the use of the recoupling formula which regulates the replacement
of the double anticommutator $\{\{ A_{j_1} , B_{j_2} \}_l , C_{j_3}
\}_L$ by a linear combination of double anticommutators with
cyclically permuted entries plus a linear combination of
{\it double commutators}. That part of each individual result
of procedure {\it a}) and of procedure {\it b}) which is due to
the occurrence of such double commutators is called the
``subleading part'' of the result of the respective procedure.
Its complement is called the ``leading part'' of the result
of the respective procedure.

An explicit factor $\hq^2$ could be
extracted from each subleading part in keeping with the
regular scaling behaviour of the remainder when $\hq$ tends to zero.
The replacement of commutators
$[ \cdot , \cdot ]_j$ by ``adapted'' commutators $\frac1{\hq}
\, [ \cdot , \cdot ]_j$ is all that would be required.

While there does not exist any difficulty concerning the execution
of procedure {\it b}) (it turns out that the results are unaffected
by special choices of the functions $\xi$ and $\vartheta$), the
execution of procedure {\it a}) requires some reflection. For carrying
out the latter procedure, especially when evaluating the commutators
$[ \XZQ{0} , \XZQ{0} ]_L$, $[ \ZZQ{0} , \ZZQ{0} ]_L$ and 
$[ \ZZQ{0} , \XZQ{0} ]_L$, two strategies come to mind.
\\
{\it Strategy no 1} (explained for the cases DYN1, $\dots$, DYN3)
\\
In the leading part, commutator by
commutator, at the expense of contributions from terms 
$\{[\XZQ{0},(\PE{1}{0}^2)_2]_L,\BNE{0}\}_L$,
$\{[\XZQ{0},(\PE{1}{0}^2)_2]_k,\TZ{0}\}_L$
and
$\{[\XZQ{0},(\PE{1}{0}^2)_0]_2,\TZ{0}\}_L$ \linebreak
($\{[\ZZQ{0},\PE{1}{0} ]_k,\SE{0}\}_L$ and 
$\{[\ZZQ{0},\PE{1}{0} ]_k,\SZ0\}_L$), move the irreducible
tensor operator $\XZQ{0}$ ($\ZZQ{0}$) as a whole into commutator
brackets pairing it with the generators $\BNE{0}$ and $\TZ{0}$
($\SE{0}$ and $\SZ{0}$), originally contained inside the irreducible
tensor operator $\ZZQ{0}$ ($\XZQ{0}$). Left with a linear combination
of terms $\{[\XZQ0,\BNE0]_2,(\PE10^2)_2\}_L$, 
$\{[\XZQ0,\TZ0]_k,(\PE10^2)_2\}_L$ and $\{[\XZQ0,\TZ0]_L,(\PE10^2)_0\}_L$ 
($\{[\ZZQ0,\SE0]_k,\PE10\}_L$ and $\{[\ZZQ0,\SZ0]_k,\PE10\}_L$) apply
the Leibniz rule, i.e. the recoupling formula for a commutator
containing an anticommutator. The outcome is a sum of two types of
linear combinations: The type-one linear combination consists of terms
in which $\BNE0$ and $\TZ0$ ($\SE0$ and $\SZ0$) are paired with
$\PE10$ ($(\PE10^2)_2$ or $(\PE10^2)_0$) inside commutators, these
commutators being paired with $\SE0$ or $\SZ0$ ($\BNE0$ or $\TZ0$)
inside anticommutators. \\
The type-two linear combination consists of
terms in which $\BNE0$ and $\TZ0$ ($\SE0$ and $\SZ0$) are paired with
$\SE0$ or $\SZ0$ ($\BNE0$ or $\TZ0$) inside commutators, these
commutators being paired with $\PE10$ ($(\PE10^2)_2$ or $(\PE10^2)_0$)
inside anticommutators.\\
With the help of Biedenharn's and Elliot's sum rule for $6-j$ symbols
(cf. e.g. Ref. \cite{RBMW 59}) show that in the leading part the type-one
linear combination can be arranged \linebreak  in 
a linear combination of terms
$\{[\ZZQ0,\PE10]_k,\SE0\}_L$ and 
$\{[\ZZQ0,\PE10]_k,\SZ0\}_L$,\linebreak[4]
($\{[\XZQ0,(\PE10^2)_2]_L,\BNE0\}_L$, 
$\{[\XZQ0,(\PE10^2)_2]_k,\TZ0\}_L$
and $\{[\XZQ0,(\PE10^2)_0]_2,\TZ0\}_L$).
Making use of i) the generalized commutation relations and ii) the
particular combinations of commutators $[\ZZQ0,\XZQ0]_L$ for
DYN1, \dots, DYN3, show for each separate DYN that in its leading part
the net contribution from the type-two combinations can be written as
a sum of anticommutators none of which containing more than one generator from
the collection of generators $\BNE0$, $\TZ0$, $\SE0$, $\SZ0$.\\
Note: The execution of procedure a) for the leading part does not
require any specification of the commutators $[\BNE0,\PE10]_1$,
\dots, $[\SZ0,\PE10]_L$ other than that of the commutators
$[\ZZQ0,\PE10]_L$ and $[\XZQ0,\PE10]_L$.
\\[1ex]
{\it Strategy no 2}
\hspace{1em}
Rewrite the generalized commutation relations for branch [0] in terms
of  $\se$, $\XZQ{0}$;
$\te$, $\ZZQ{0}$ and insert the resulting version of the commutation
relations into DYN0, \dots, DYN3. Check
that in the leading part of the DYNs after this insertion no commutators
are left which cannot be cast into the form $[ \XZQ{0} , \PE{1}{0} ]_L$
and $[ \ZZQ{0} , \PE{1}{0} ]_L$. Moreover, after evaluation of these
commutators the leading parts can be written as sums of anticommutators none
of which containing the elements $\se$ and $\te$ or more than one
element from the collection $\XZQ0$, $\ZZQ0$.

\vspace{5mm}
We observe that in accordance with the correspondence principle
for each individual DYN the leading part of the
result of the two procedures {\it a}) (independent of the strategy
employed) and {\it b}) vanishes in both cases. All -- if anything
-- that survives is the subleading part of the respective result.
{\it Pars pro toto} we quote the surviving $\equiv$ subleading
parts of the result of procedure {\it a}) and of procedure {\it b})
applied to DYN0, these parts being denoted by dyn0a and dyn0b,
respectively. (As a second example, we quote the analogous parts
dyn1a and dyn1b related to DYN1 in the appendix. While the expressions
for the analogous parts dyn2a and dyn3a related to DYN2 and DYN3,
respectively, are too lengthy to be quoted in this article at all,
the expressions for the corresponding parts dyn2b and dyn3b are
sufficently simple. They too are reproduced in the appendix.)

In order to arrive at the expressions for dyn$\cdot$a (quoted and
unquoted) we had to make
extensive use of the formal properties of commutator brackets, of the
vanishing of the commutator $ [  \PE{1}{0} , \PE{1}{0}  ]_1$, and of
special derivates of the generalized commutation relations for branch [0].
These derivates are obtained as follows: We insert the expressions of the
original generators in terms
of $\se$, $\XZQ{0}$; $\te$, $\ZZQ{0}$ and  commute separately the
two sides of each generalized commutation relation twice with the
components of the irreducible tensor variable $\PE{1}{0}$. Subsequently,
we equate the corresponding results:
\[   [[ (\text{l.h.s.}) , \PE{1}{0} ]_l , \PE{1}{0} ]_L =
     [[ (\text{r.h.s.}) , \PE{1}{0} ]_l , \PE{1}{0} ]_L  \]
and work out the implications of these equations. As for the commutators
\[   [  \mathscr{C}_p , \PE{1}{0} ]_L  \qquad  \text{and}\qquad
     [  \mathscr{K}_q , \PE{1}{0} ]_L , \qquad \qquad p,q = 0,1,2  \]
turning up in the course of the act, we prefer to postpone the
injection of the previous specifications for $ \CZ$, $ \KE$ and
$ \KZ$, (and their full implications) to a later stage of the
analysis. Instead, for the time being, we work under the hypothesis
that, when evaluating the above commutators, appropriate powers of
the tensor variable $\PE{1}{0}$ completely take care of the
non-trivial $O(3)$ transformation properties of the respective
results, these powers being accompanied by appropriate real
functions $ \bar \sigma$, $ \tau_{01}$, $ \tau_{11}$ and $ \tau_{23}$
of the $O(3)$--scalar observables $ \mn$, $ \pominus^2$ and $\PN{1}{0}$.
The functions $ \bar \sigma$ and $ \tau_{23}$, having mass dimension
0, can be thought of as functions of the quotients
$ \big( \frac{ \poe -  \poz}{ \mn} \big)^2$ and $ \frac{\PN{1}{0} }{ \mnq}$,
while the functions $ \tau_{01}$ and $ \tau_{11}$ have mass dimension
2.
\\
In formulae:
\begin{equation*}
\begin{array}{lcl}
    [  \CN , \PE{1}{0} ]_1 &=& 0   \,, \\[1ex]
    [  \CE , \PE{1}{0} ]_L &=&  \sqrt{3 }  \; [  \CZ , \PE{1}{0} ]_L =
     \delta_{L,2 }  \,  \frac1{i }  \,  \hq^2  \, \PM{2}{1}{0}  \,
     \pominus  \,  \bar \sigma  \,, \\[1ex]
[ \KN , \PE{1}{0} ]_1 &=&  \hq^2 \, \PE{1}{0} \, \tau_{01},
\\[1ex]
[ \KE , \PE{1}{0} ]_L &=&  \delta_{L,1} \, \hq^2 \, \PE{1}{0} \,\tau_{11},
\\[1ex]
[ \KZ , \PE{1}{0} ]_L &=&  \hq^2 \; \left\{  \begin{array}{lcl}
      \frac1{\sqrt{5}} \: \PE{1}{0} \: \big[ 2 \, \tau_{01} - 3 \, \big(
      \frac{\tau_{11}}{\sqrt{3}} \big) \big]  \qquad
        &  \text{for} & L=1  \\
      0 &  \text{for} & L=2  \\
      \PM{3}{1}{0} \,\tau_{23} &  \text{for} & L=3 \,.
      \end{array} \right.
\end{array}
\end{equation*}
In this way we obtain the following preliminary expression for dyn0a
(and for the other dyn$\cdot$a)

%
\begin{equation*}
\begin{array}{lcl}
\multicolumn{3}{l}{\text{dyn0a =}}\\
&&
\frac1{\hq} \; \Big\langle
\frac{25}{7\,{\sqrt{3}}} \: \{[\KZ,\KZ]_{3},\PM{2}{1}{0}\}_{2} -
5 \, {\sqrt{\frac{5}{42}}} \;
\{[\KZ,\KZ]_{3},\frac{\PM{4}{1}{0}}{\PN{1}{0}}\}_{2} + \\
&&
4 \, {\sqrt{\frac{5}{21}}} \; \{[\KZ,\KN]_{2},\PN{1}{0}\}_{2} -
\frac{20}{7} \, {\sqrt{\frac{5}{3}}} \;
\{[\KZ,\KN]_{2},\PM{2}{1}{0}\}_{2} + \\
&&
\frac{5}{{\sqrt{21}}} \:
\{[\KZ,\KN]_{2},\frac{\PM{4}{1}{0}}{\PN{1}{0}}\}_{2}
\Big\rangle -
\frac1{\hq} \;
\Big\langle   \KZ  \leftrightarrow  \CZ , \;
              \KN  \leftrightarrow  \CN      \Big\rangle +  \\
&&
\frac{8}{{\sqrt{35}}} \: \hq \,
\{\KZ, \big\langle
{\sqrt{3}} \: {{\tau }_{01}} + {{\tau }_{11}} -
{\sqrt{3}} \: {{\tau }_{23}} \, \PN{1}{0} \big\rangle \:
\PN{1}{0}\}_{2} - \\
&&
\frac{8}{7} \, {\sqrt{5}} \; \hq \,
\{\KZ, \big\langle
{\sqrt{3}} \: {{\tau }_{01}} + {{\tau }_{11}} -
{\sqrt{3}} \: {{\tau }_{23}} \, \PN{1}{0} \big\rangle \:
\PM{2}{1}{0}\}_{2} + \\
&&
\frac{2}{{\sqrt{7}}} \: \hq \,
\{\KZ, \big\langle
{\sqrt{3}} \: {{\tau }_{01}} + {{\tau }_{11}} -
{\sqrt{3}} \: {{\tau }_{23}} \, \PN{1}{0} \big\rangle \:
\frac{\PM{4}{1}{0}}{\PN{1}{0}}\}_{2} + 
\\
&&
i \, 72 \, {\sqrt{\frac{2}{7}}} \; \hq \,
\{\CZ,\pominus\,\PN{1}{0}\,\PE{1}{0}\}_{2} -
i \, 12 \, {\sqrt{5}} \; \hq \,
\{\CZ,\pominus\,\PM{3}{1}{0}\}_{2} + \\
&&
\hq^2  \: \bigg(
192 \, {\sqrt{\frac{30}{7}}} \;
\{\JE{0},\frac{\PM{2}{1}{0}}{\PN{1}{0}}\}_{2} \:
\Big\langle  \big( \potimes - {\sqrt{3}} \, \PN{1}{0} \big) \
{\PN{1}{0}}^2  \Big\rangle - \\
&&
i \, 24 \,{\sqrt{\frac{30}{7}}} \;
\{\AE,\frac{\PM{2}{1}{0}}{\PN{1}{0}}\}_{2} \:
\Big\langle  \big(  7 \, \sqrt{3} \, \potimes + 8 \, \PN{1}{0} \big)
\, \PN{1}{0}  \Big\rangle + \\
&&
i \, 192 \, {\sqrt{\frac{30}{7}}} \;
\{\BE,\frac{\PM{2}{1}{0}}{\PN{1}{0}}\}_{2} \:
\Big\langle  \pominus \, {\PN{1}{0}}^2  \Big\rangle
\bigg)  \,,
\end{array}
\end{equation*}
\beas
\lefteqn{\text{dyn0b } = }  \\ &&
\text{$\hq^2$ times the content of the bracket $\big(  \,  \big)$
of the preceeding expression for dyn0a. }
\eeas
Now we make use of the previous specifications. We start with those for
$ \KE$ and $ \KZ$. By themselves, by Jacobi identities, and by the
vanishing of the commutator $ [  \PE{1}{0} , \PE{1}{0}  ]_1$, the
specifications imply 1.) the correctness of the previous hypothesis
for the commutators $ [  \mathscr{K }_q , \PE{1}{0}  ]_L$ and 2.)
the explicit form of the functions $ \tau_{01}$, $ \tau_{11}$ and
$ \tau_{23}$:
\[  \tau_{01} = - {\textstyle\frac{8}{3} }\, {\sqrt{5}} \; \potimes  ,\qquad
    \tau_{11} = -4 \, {\textstyle{\sqrt{\frac{5}{3}}}} \;  \potimes  ,\qquad
    \tau_{23} = 0  \:.  \]
Next we turn to the specification for $ \CZ$ in terms of $ \CN$ and
$ \CE$. Working with the still unverified part of the hypothesis and
inserting the complete expressions for  the commutators $ [  \mathscr{K }_q ,
\PE{1}{0}  ]_L$ plus the specification for $ \CZ$ into a scalar derivate
of the  generalized commutation relations from the second half of the
previous list, we find the explicit expression for $ \CN$:
\[  \CN  =  - 4 \, {{\textstyle\sqrt{\frac{5}{3}}}}\;  \hq \; \{\BE,\PE{1}{0}\}_{0}  \,.\]
As a corrolary we obtain: $ [ \CN , \PE{1}{0}  ]_1 = 0$. Without
further assistance of the derivates, this together with the previous
specification for $ \CZ$, the Jacobi identities and the vanishing of the
commutator $ [  \PE{1}{0} , \PE{1}{0}  ]_1$ implies
1.) the self-consistency of the previous hypothesis for the commutators
$ [  \mathscr{C }_p , \PE{1}{0}  ]_L$ and 2.) the explicit form
of the ``function'' $ \bar \sigma$:
\[  \bar \sigma = 3  \: {\textstyle\sqrt{\frac65 }}  \,. \]

\vspace{5mm }
If we insert the explicit forms of $ \tau_{01}$, $ \tau_{11}$ and
$ \tau_{23}$ and the specifications for $ \CZ$, $ \KZ$ and $ \KE$
into the above formula for dyn0a, the contribution of all terms
preceeding those inside the bracket $( \: )$ at the end vanishes.
Thereby we have established the equality of dyn0a and dyn0b
which is part of our consistency check.

The numerical value $3  \: \sqrt{\frac65 }$ of $ \bar \sigma$ and
additional assistance of the derivates of the generalized
commutation relations is required in order to show that the remaining
dyn$\cdot$a's are equal to the respective dyn$\cdot$b's. Required are
expressions for $ [ \CE, \AE ]_1$, $ [ \CE, \AE ]_2$,
$ [ \te, \pominus ]_1$ and
$ [ \KN, \CE ]_1$. To illustrate the structure of these
expressions we quote the ones for $ [ \te, \pominus ]_1$ and
$[ \KN, \CE ]_1$ in the appendix, the ones for $[ \CE, \AE ]_1$
and $[ \CE, \AE ]_2$ being equally lengthy if not even more so.
In all cases the question whether the quantal composition laws
$\XZQ{0} = \dots$ and $\ZZQ{0} = \dots$ are in agreement with the
generalized commutation relations has been answered affirmatively.


\section*{Conclusions}

In the same way as selection rules limit the possible transition
processes in atomic and elementary particle physics, (non-additive)
composition laws in string physics restrict the possible outcome of
the merging and splitting of strings. Within the (3+1)--dimensional
Nambu--Goto theory we considered the subsector of bosonic closed
strings in which all interactions proceed via string vertices,
\ie\ via merging and splitting.

In the classical setting two such tensorial non-additive composition laws
had already been established \cite{Grosser}. In the first place,
they had furnished numerical relations among the observables
related to the three separate branches of the string vertex.
On top of that, their r\^ole as generators for further non-additive
composition laws by means of Poisson bracket operations had led to
an attempt to count the number of new composition laws obtainable
from them by such induction.

Based on this classical construction, we have explored the viability
of suitably adapted versions of these composition laws in the
corresponding quantum theory. In order to settle this issue,
we looked for rigorous tests of the composition laws as relations
in an associative graded algebra, its non-commutativity being
curbed by the generalized commutation relations for the quantum
observables related to the three separate branches of the vertex.

Apart from the obvious hermiticity, covariance and exchange symmetry
requirements, the crucial tests entered in the form of two special
types of constraints: 1.) a ``kinematical constraint'' in the context
of {\it manifest} $O(3)$--covariance and 2.) several ``dynamical
constraints'' probing the consistency of the quantized composition
laws with the generalized commutation relations. The quantized composition
laws passed all these tests unforcedly. Thus they can be regarded
as a well-established structural element of the quantum theory of
closed Nambu--Goto strings.

Now we are in a position to evaluate the additional mixed commutators
of $\XZQ{0}$ and $\ZZQ{0}$ on the one hand and of the generators of
the algebras $\ghh^{[k]-}$, $k=1,2$, on the other hand --- as far as
the latter algebras and our explicit knowledge of them permit. The
as yet undetermined functions $ \xi$ and $ \vartheta$ drop out of
the pertinent calculations.

Further, due to the infinite dimension of the algebra of observables
and in striking contrast to the Huygens-Newtonian laws of impact, by
commutator induction we are in a position to produce more 
and more new quantized non-additive composition laws (which now in general
do involve the functions $ \xi$ and $ \vartheta$). As we have already
established, in this way we obtain all new non-additive composition laws
(of the restricted type) within the $P$--linear span of $\GVH^{[k]2}$.

Certainly, the functions $ \xi$ and $ \vartheta$ occurring on the r.h.s.\
of the composition laws $\XZQ{0} = \dots$ and $\ZZQ{0} = \dots$,
respectively, could be narrowed down by requiring i) covariance of the
composition laws under Lorentz transformations, and ii) observance of
the {\em formal} crossing symmetry (see above).

From the second requirement one could derive two linear inhomogeneous
functional equations relating the functions $\xi$ and $\vartheta$ with
arguments
$\left(\frac{\PO1-\PO2}{\mn}\right)^2$, $\frac{(\PE10^2)_0}{(\mn)^2}$
to the same functions with arguments
$\left(\frac{\PZ1+\PZ2}{\mz}\right)^2$, $\frac{(\PE12^2)_0}{(\mz)^2}$.
The inhomogeneities would be calculable from the double commutators
arising in the course of the rearrangement of the composition laws 
$\XZQ0=\ldots$, $\ZZQ0=\ldots$ (for the real process: branch [1]
merging with branch [2] to form branch [0]) into the composition laws 
$\XZQ2=\ldots$, $\ZZQ2=\ldots$ \linebreak 
(for the non-real process: branch [1]
merging with crossed branch [0] to form crossed branch [2]).\\
This option will be pursued once arguments will have been presented 
that the {\em formal} crossing symmetry 
must also be observed in quantum theory.

\vspace{5mm}
\noindent
Responsibilities: K.P.\ for the analytic line of reasoning and for the
formulation of the manuscript, M.T.\ for the symbolic computations.


\section*{Appendix}
\begin{equation*}
\begin{array}{lcl}
\multicolumn{3}{l}{\text{dyn1a}= }\\
&&
\frac1{\hq} \; \Big\langle
i \, \frac{12}{7 \, {\sqrt{5}}} \:
\{[\KZ,\CZ]_{3},\frac{\PM{3}{1}{0}}{\PN{1}{0}}\}_{0} -
i \, \frac{32}{35} \, {\sqrt{\frac{2}{7}}} \;
\{[\KZ,\CZ]_{1},\PE{1}{0}\}_{0} -
i \, \frac{6}{7} \, {\sqrt{\frac{2}{5}}} \;
\{[\KZ,\CE]_{3},\frac{\PM{3}{1}{0}}{\PN{1}{0}}\}_{0} + \\[1ex]
&&
i \, \frac{4}{35} \, {\sqrt{\frac{2}{21}}} \;
\{[\KZ,\CE]_{1},\PE{1}{0}\}_{0} -
i \, \frac{22}{21} \, {\sqrt{\frac{2}{5}}} \;
\{[\KE,\CZ]_{3},\frac{\PM{3}{1}{0}}{\PN{1}{0}}\}_{0} -
i \, \frac{4}{35} \, {\sqrt{\frac{2}{21}}} \;
\{[\KE,\CZ]_{1},\PE{1}{0}\}_{0} - \\[1ex]
&&
i \, \frac{8}{7} \, {\sqrt{\frac{2}{35}}} \;
\{[\KE,\CE]_{1},\PE{1}{0}\}_{0} +
i \, 4 \, {\sqrt{\frac{2}{105}}} \;
\{[\KE,\CN]_{1},\PE{1}{0}\}_{0} -
i \, 4 \, {\sqrt{\frac{2}{105}}} \;
\{[\KN,\CE]_{1},\PE{1}{0}\}_{0}
\Big\rangle + \\[1ex]
&&
i \, \frac{128}{7} \, {\sqrt{\frac{2}{105}}} \;
\{[\KZ,\BE]_{2},\PM{2}{1}{0}\}_{0} -
i \, \frac{128}{35} \, {\sqrt{\frac{2}{7}}} \;
\{[\KE,\BE]_{0},\PN{1}{0}\}_{0} - \\[1ex]
&&
\frac{358}{21\,{\sqrt{35}}} \:
\{[\CZ,\AE]_{1},\PE{1}{0}\}_{0} -
\frac{922}{35\,{\sqrt{21}}} \:
\{[\CE,\AE]_{1},\PE{1}{0}\}_{0} + \\[1ex]
&&
\frac{4}{945\,{\sqrt{7}}} \, \hq \, \{\KZ, \big\langle
261\,{\sqrt{5}}\;\pominus - 370 \,{\sqrt{6}}\;\pominus \,\bar\sigma
\big\rangle  \,\PM{2}{1}{0}\}_{0} - \\[1ex]
&&
\frac{8}{945\,{\sqrt{7}}} \, \hq \, \{\KN, \big\langle
1125\,\pominus - 2 \, {\sqrt{30}} \; \pominus \,\bar\sigma \big\rangle
\,\PN{1}{0}\}_{0} - \\[1ex]
&&
\frac{32}{7} \, \hq \, {\sqrt{\frac{2}{105}}} \;
\{[\te,\pominus \,\bar\sigma ]_{1},\PN{1}{0}\,\PE{1}{0}\}_{0} +
i \, \frac{2}{105} \, {\sqrt{\frac{2}{7}}} \; \hq \, \{\CE, \big\langle
44\,{\sqrt{15}} \; {{\tau }_{01}} +
20\,{\sqrt{5}} \; {{\tau }_{11}} + \\[1ex]
&&
4\,{\sqrt{15}} \; {{\tau }_{23}}\,\PN{1}{0} +
273\,{\sqrt{3}} \; \potimes +
216\,\PN{1}{0}  \;
\big\rangle \, \PE{1}{0}\}_{0}   \,,  \\[2ex]
\multicolumn{3}{l}{\text{dyn1b}= 0}
\end{array}
\end{equation*}
%
%
%
\begin{equation*}
\begin{array}{lcl}
\multicolumn{3}{l}{\text{dyn2b =}}\\
%
&&    \hq^2  \: \bigg(
    - 48 \, {\sqrt{\frac{14}{5}}} \;
    \{\JE{0},\frac{\PM{2}{1}{0}}{\PN{1}{0}}\}_{2} \,
    \Big\langle \pominus\,{\PN{1}{0}}^3 \Big\rangle - \\
&& \qquad
    i \, 144 \, {\sqrt{\frac{42}{5}}} \;
    \{\AE,\frac{\PM{2}{1}{0}}{\PN{1}{0}}\}_{2} \,
    \Big\langle \pominus\, {\PN{1}{0}}^2 \Big\rangle - \\
&& \qquad
    i \, 48 \, {\sqrt{\frac{42}{5}}} \;
    \{\BE,\frac{\PM{2}{1}{0}}{\PN{1}{0}}\}_{2} \,
    \Big\langle \big( \potimes + \sqrt{3} \, \PN{1}{0} \big) \,
    {\PN{1}{0}}^2 \Big\rangle  \bigg)  \:,  
\qquad\qquad\qquad\qquad
\\[2ex]
%
%
%
\multicolumn{3}{l}{
\text{dyn3b =}}\\
%
&&    \hq^2  \: \bigg(
    - 28704 \, {\sqrt{3}} \;
    \{\JE{0},\frac{\PM{4}{1}{0}}{{\PN{1}{0}}^2}\}_{4} \,
    \Big\langle \pominus\, {\PN{1}{0}}^3 \Big\rangle - \\[2ex]
&& \qquad
    i \, 107699 \,
    \{\AE,\frac{\PM{4}{1}{0}}{{\PN{1}{0}}^2}\}_{4} \,
    \Big\langle \pominus\, {\PN{1}{0}}^2 \Big\rangle + \\
&& \qquad
    i \, 96 \,
    \{\BE,\frac{\PM{4}{1}{0}}{{\PN{1}{0}}^2}\}_{4} \,
    \Big\langle \big( 19\,\potimes - 373\,{\sqrt{3}}\,\PN{1}{0} \big) \,
    {\PN{1}{0}}^2 \Big\rangle \bigg)  \: ,
\qquad\qquad\qquad\qquad
%
%
%
\\[2ex]
\multicolumn{3}{l}{
[ \te , \pominus ]_1 = 
- i \, \frac{64}{63} \, {\sqrt{2}} \; \CE -
i \, \frac{5}{63} \, {\sqrt{10}} \;
\{\CE,\frac{\PM{2}{1}{0}}{\PN{1}{0}}\}_{1} -
i \, \frac{4}{7} \, {\sqrt{30}} \; \{\BE,\PE{1}{0}\}_{1}    \,,}
\\[2ex]
%
%
%
\multicolumn{3}{l}{
    [ \KN,\CE ]_1  =
    i \, {\sqrt{\frac{6}{5}}} \;
    \{\KN,\pominus\,\PE{1}{0}\}_{1} +
    \frac{44}{21} \, {\sqrt{\frac{5}{3}}} \;
    \{\CE,\PN{1}{0}\}_{1} +
    \frac{100}{21\,{\sqrt{3}}} \:
    \{\CE,\PM{2}{1}{0}\}_{1} - }
\\[1ex]
&&
    \frac{8}{3} \, {\sqrt{5}} \; \{\CE,\potimes\}_{1} -
    i \, \frac{40}{3} \, \{\JE{0},\PE{1}{0}\}_{1} \,
    \big\langle  \pominus \, \PN{1}{0} \big\rangle  + \\[1ex]
&&
    \frac{56}{3\,{\sqrt{3}}} \: \{\AE,\PE{1}{0}\}_{1} \,
    \big\langle  \pominus  \big\rangle  -
    \frac{16}{63} \, {\sqrt{3}} \; \{\BE,\PE{1}{0}\}_{1} \,
    \big\langle  35 \, \potimes + 4 \, {\sqrt{3}} \, \PN{1}{0} \big\rangle \,.
\end{array}
\end{equation*}


\end{document}